\documentclass[manuscript]{aastex}

\slugcomment{Not to appear in Nonlearned J., 45.}

\shorttitle{Identification of Potential Sites}
\shortauthors{Pinz\'on et al.}

\begin{document}


\title{Identification of Potential Sites for Astronomical Observations\\
in Northern South-America}


\author{G. Pinz\'on\altaffilmark{1}, D. Gonz\'alez\altaffilmark{1}}
\affil{Observatorio Astron\'omico, Facultad de Ciencias, Universidad Nacional de Colombia, Carrera 45 No. 26-85,  Bogot\'a, Colombia;  gapinzone@unal.edu.co, dagonzalezdi@unal.edu.co.}

\and

\author{J. Hern\'andez\altaffilmark{2}}
\affil{Centro de Investigaciones de Astronom\'ia (CIDA), Apdo. Postal 264, M\'erida 5101-A, Venezuela; jesush@cida.ve.}




\begin{abstract}

In this study we describe an innovative method to determine potential sites for optical and infrared astronomical observations in the Andes region of northern South America. The method computes the Clear sky fraction (CSF) from Geostationary Observational Environmental Satellite (GOES) data for the years 2008-12 through a comparison with temperatures obtained from long-term records of weather stations and atmospheric temperature profiles from radiosonde. Criteria for sky clearance were established for two infrared GOES channels in order to determine potential sites in the Andes region of northern South-America.  The method was validated using the reported observed hours at \textit{Observatorio Nacional de Llano del Hato} in Venezuela. Separate CSF percentages were computed for dry and rainy seasons for both, photometric and spectroscopic night qualities. Twelve sites with five year averages of CSF for spectroscopic nights larger than 30\% during the dry seasons were found to be suitable for astronomical 
observations. The best site with (220$\pm$42) spectroscopic clear nights per year is located in the Andes of Venezuela (70$^{\circ}$28'48"W, 9$^{\circ}$5'60"N) at an altitude of 3480 meters.  Lower quality regions were found in \textit{Sierra Nevada de Santamarta} and \textit{Serran\'ia del Perij\'a} with (126$\pm$34) and (111$\pm$27) clear nights per year, respectively. Sites over the Andes are identified in \textit{Norte de Santander} with (107$\pm$23) and in the north-east part of \textit{Boyac\'a} with a mean of (94$\pm$13) clear nights per year. Two sites at low latitude located in Ecuador with more than 100 clear nights per year and with similar seasonal CSF percentages were also identified. Five year evolution suggest a possible correlation between the lowest percentages observed during the rainy seasons of 2010 and 2011 with positive values of the Southern Oscillation Index. 
\end{abstract}


\keywords{Astronomical Instrumentation. }



\section{Introduction}

The use of images of the earth in different wavelengths from infrared cameras on board of geostationary satellites is a powerful tool for the identification of new potential sites for astronomical observations in optical, infrared and submillimeter wavelengths (Erasmus \& Sarazin 2002; Cavazzani et al. 2012; Cavazzani \& Zitelli 2013). Sites in remote areas with complex relief and scarse or null meteorological records can be identified by previous computation of the Clear Sky Fraction (CSF). As the cloud covering is a function of the height of the atmosphere, images from satellite are useful due to the detectors on board measure the emissivity of the atmosphere at different layers. 
 
Several studies for the determination of the CSF are found in the literature. Soden \& Bretherton (1993) conducted an analysis of upper tropospheric humidity and clouds using GOES observations, finding a correlation between humidity and cloud cover at high altitudes. Erasmus \& Sarazin (2002) were among the first to sucesfully use long-term satellite data for CSF computations for a wide range of altitudes. Erasmus \& van Rooyen (2006) and Della Valle et al. (2010) reproduced with high precision the CSF records of ground data at la Palma with a model based on satellite data. Cavazzani et al. (2011,2012) found percentages of clear nights at five astronomical observatories around the world using long-term GOES data. Hidayat et al. (2012) used 15 years of GOES infrared images in 6.7 and 10.7$\mu$m bands  to compute the percentage of clear sky over a wide region in Indonesia. Recently, Mar\'in et al. (2015) computed the CSF at high altitude sites by using a new clearance algorithm based on global meteorological 
models. CSF values were then used for precipitable water vapour estimations at two astronomical sites in the north of Chile.

Even though most of the methods are based on the comparison of radiances, fluxes or temperatures at each pixel of the satellite image with threshold values  previously defined there is no standard methodology for CSF computations.  Cavazzani \& Zitelli (2013) used monthly thresholds defined as the maximum radiance at clear nights along the month. Thresholds may also be defined by using an atmospheric model \citep{H12}. In this study, we introduce a novel methodology involving the use of monthly thresholds of temperature at different altitudes from the surface up to the upper troposphere. Our thresholds definitions were based on the altitude of the terrain, long-term records of weather stations and radiosonde data archives.

The thresholds definitions are based on monthly mean records of both, air temperature at the site and vertical temperature profile between 8 and 10 km of altitude which make more precise the predictions for clear covering than those involving atmospheric models for thresholds computations.  Our methodology was applied in northern South America over the tropical region covering Colombia, western Venezuela and northern Ecuador. Even though the zone is  far from being interesting for infrared and submilimeter astronomy  due to the requirement of excellent weather conditions, many universities and institutes from those countries are planning to develope optical observatories. The countries are located within the intertropical convergence zone (ITCZ) showing complex topography and with a climate governed by a bimodal annual cycle which results of the double passage  of the ITCZ for the region (Poveda et al. 2011). The Andes affect the climate by blocking zonalflows, influencing regional wind patterns and 
precipitation and hosting diverse micro-climates. The main goal of this study is to seek prospective locations with middle quality CSF values over the Andes of those countries. 

The paper is organized as follows. In Section 2 we describe the satellite data used in this study. Threshold temperatures calculations are presented in Sections 3 and 4. The methodology and validation are discussed in Sections 5 and 6 while in Sections 7 and 8 we present the results and the selected sites.

\section{Satellite based data}


The Geostationary Operational Enviromental Satellites GOES 12 and 13 are equiped with instrumentation capable of capture images in several infrared bands. For our purposes we concentrate our attention on B3 and B4 bands centered at 6.7 and 10.7$\mu$m, respectively. Images in B4 are strongly affected by emission and absorption of water vapour found at middle altitudes ($\sim$4 km) whereas layers of the upper trophosphere between 8 and 10 km are sensed by images in the B3 band. Other bands as those centered in 13.3$\mu$m provide information about fog and very low atmosphere characteristics. 

Since the data extends for decades these satellites are commonly used for long-term studies. Temporal domaine for this study begins on the 1th of January 2008 and ends on the 31st of December 2012 with a few missing data lower than 7\% in the worst cases. GOES missing data has different causes: technical troubles in GOES-12 (from December 15th 2008 to January 6th 2009); when GOES-13 replaced GOES-12 (from April 14th to 28th 2010); and due to anomalies in the operational system (from September 23th to October 18th 2012). We assumed the clear sky fraction as null for the days with GOES missing data.
  
GOES 12-13 images are freely available through an electronic library of NOAA\footnote{National Oceanic and Amospheric Administration} in format GVAR (GOES Variable Format). We selected a region of the image to download covering a big part of Colombia, the western part of Venezuela and the north of Ecuador. For each night we used available observed time at the three different hours 00:15:15, 06:45:13 and 09:45:14 UTC in order to cover the entire night. GVAR is an instrumental radiance given in counts. We used the library of \textit{idl} routines  provided by the National Climatic Data Center (NCDC) for converting the counts to radiance at each pixel. Usual calibration of the images  was done  using the scaling bias and gain values available from NCDC website\footnote{http://www.ncdc.noaa.gov/} in order to get the radiance in $mW/ [m^2 sr cm^{-1}]$. 

Brightness temperatures $T^b_i$ at the instants $i=1,2,3$ of the night are computed from calibrated radiance values by assuming black body emission. Conversion to actual temperatures is done using the  relation $T_i=\alpha + \beta T^{b}_i$ where $\alpha$ and $\beta$ are conversion coefficients dependent on each band (Weinreb et al. 1997, Rossow et al. 1985). Reported values available from NCDC website, for GOES 12 satellite  are $\beta=$1.01 and $\alpha=-4.76$K for the B3 band and $\beta=$1.00 and $\alpha=-0.36$K for the B4 band. The differences between the actual and brightness values increase with decreasing temperature but usually are lower than 4\%. For instance, a brightness temperature of $10000$K corresponds to  temperatures  of $1095$K and $9999.6$K for the B3 and B4 bands, respectively. For most of cases those differences are negligible for the CSF computations. We indicate actual temperatures in B3 and B4 bands as $T_{B3,i}$ and  $T_{B4,i}$, respectively.

The spatial resolution of GOES images is 4$\times$4 km in both bands and thus a single pixel subtends only a small field of view from the ground compared with the total field of view available for an observer at ground.  As the validation of our method involves comparison with log-books of the \textit{Observatorio Nacional de Llano del Hato} in Venezuela (hereafter, CID) we degraded  the original resolution in order to make more real comparisons. Other authors have used similar or even lower resolutions. Cavazzani and Zitelli (2012) computed GOES temperatures for each site over a $1^{\circ}\times 1^{\circ}$ sub-matrix which corresponds to a linear projection of about 100$\times$100 km centered on the coordinates observatories at ground. We used a Delanuay interpolation of actual temperatures $T_{B3,i}$ and  $T_{B4,i}$ on a grid of 51$\times51$ pixels obtaining a resolution of 28$\times$18 km which we consider wide enough for validation with ground data.




\section{ Threshold temperature for middle-high clouds detection}

Detection of mid-altitude clouds at a height of 4000 meters above a pixel location of the image in B4 band and at a fixed instant during the night requires the previous evaluation of a threshold or reference temperature. In this study we compute monthly thresholds of temperature based on long-term records of weather stations located inside the region covered by the GOES B4 and B3 images.   \\

\begin{figure}
\plottwo{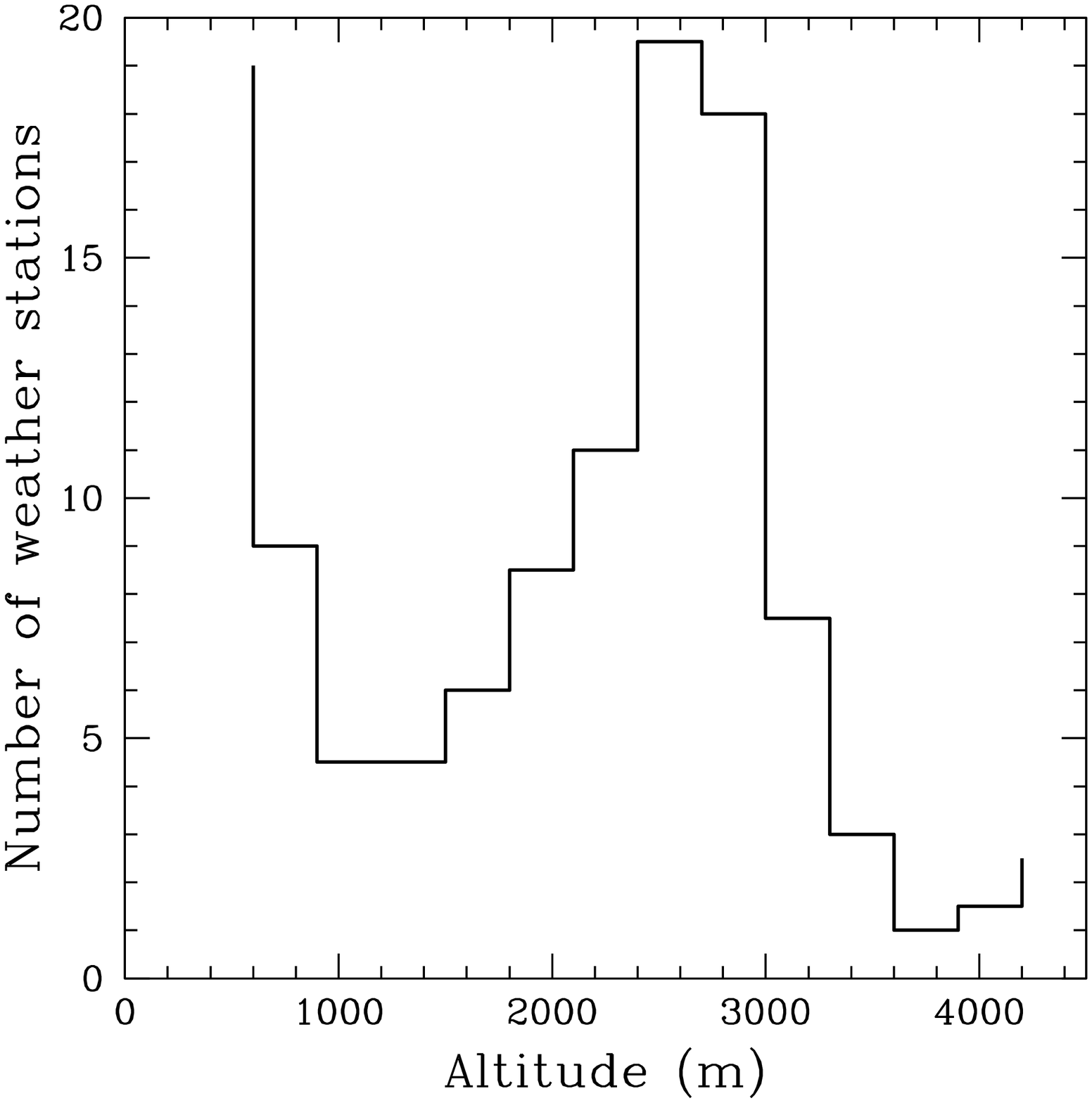}{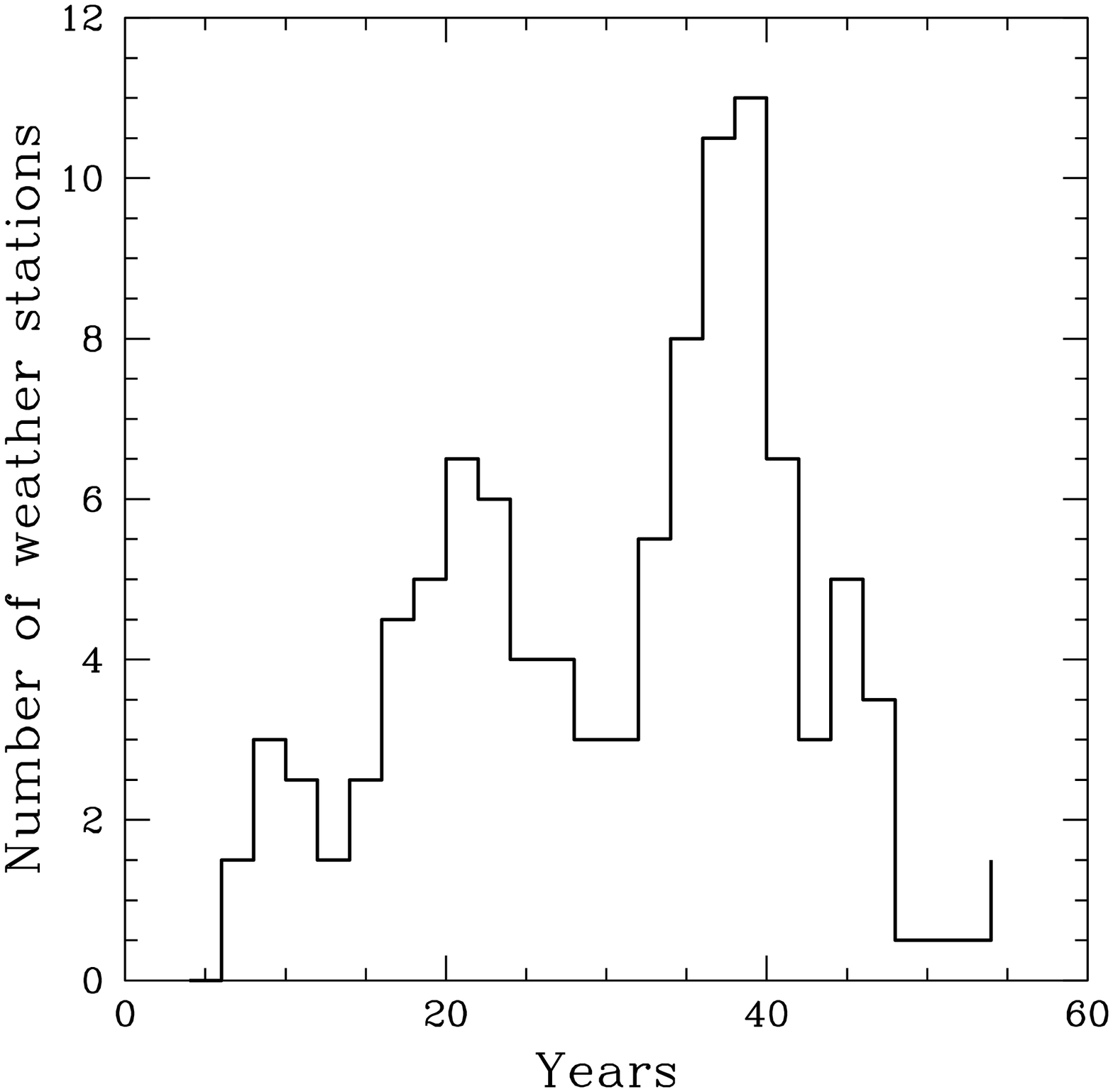}
\caption{Distribution of the IDEAM long-term records with altitude (left) and with the number of years of records (right).}\label{F1}
\end{figure}

We used long-term records of meteorological data obtained from the \textit{Instituto de Hidrolog\'ia, Meteorolog\'ia y Estudios Ambientales}, hereafter IDEAM. The IDEAM data archive comprise periods from 7 to 40 years of monthly mean values of temperature and precipitation for a set of weather stations in Colombia. The complete archive contains 400 records  at altitudes between 400 and 4700 meters above sea level covering 25\% of the country.  The archive presents a high number of gaps, especially those of temperature. We considered only long-term records of temperature with a maximum of one monthly absence obtaining the reduced number of 63 weather stations irregularly distributed in a region comprising  latitudes between 3$^{\circ}$ and 7$^{\circ}$ North, and longitudes from -76$^{\circ}$ up to -72$^{\circ}$ West.  We will refer to this region as the control region defined by the spatial distribution of the IDEAM long-term historical records. The longest records of these selected stations have more than 40 
yr as is shown in the right panel of Figure \ref{F1}. Locations are widely distributed in different altitudes with a high number of them close around Bogot\'a (BOG) ($h$=2600 meters) as shown in the left panel of the same figure.\\

\begin{figure}
\epsscale{.80}
\plotone{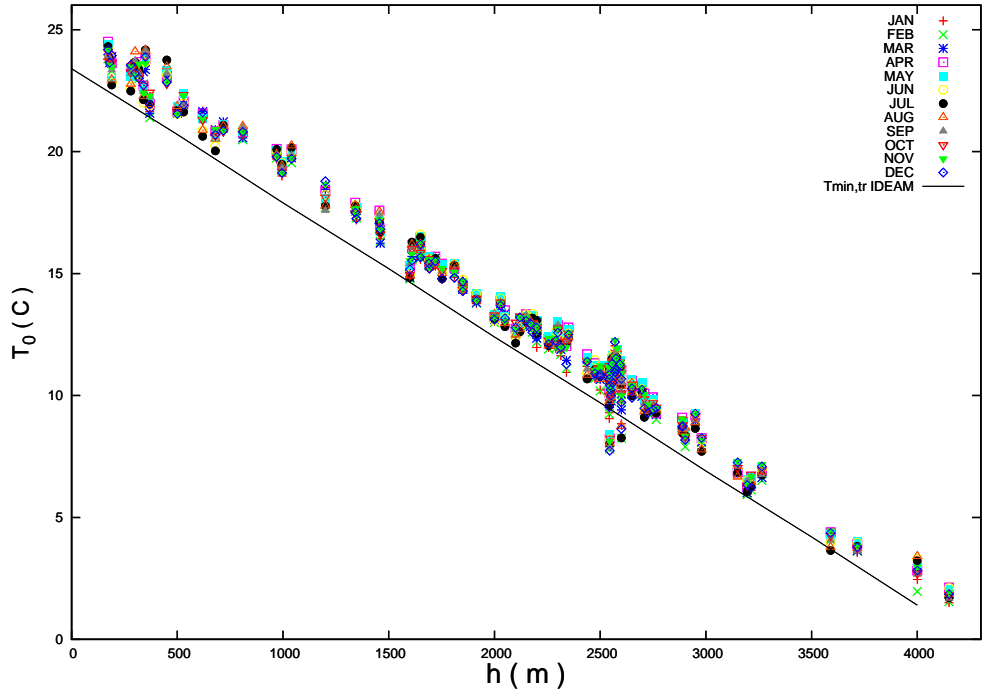}
\caption{Historical monthly thresholds of temperature at the locations of the weather stations with long-term records as a function of the altitude. Solid line represents the trend of the minima temperature obtained from short-term records: $T_{min,tr}=-5.5^{\circ}$C/km$\times h +23.42^{\circ}$C where $h$ is the altitude of the station in km.}\label{F2}
\end{figure}

The average of monthly mean records at each station for periods spanning at least 7 years is expected to be normally distributed around a central value $<T_s>$ being very likely that the monthly mean values are within $3\sigma_{s}$ of the distribution, where $\sigma_{s}$ is the standard deviation.  The coldest temperature at each station may be then defined as $T_{min}=<T_{s}>-3\sigma_{s}$. However, an additional correction term must be applied since the measurements do  not represent values during the night. We recall  that the IDEAM archive consists only of mean temperatures thus we must  correct them for nightime observations. The monthly threshold temperature at each station may therefore computed as :
\begin{equation}
T_{0}\equiv\frac{1}{2}[T_{min}-T_{min,tr}]
\label{e1}
\end{equation}
where $T_{min,tr}$ is the trend of the minima records of temperature given by short-term records of weather stations indicated with a solid line in Figure \ref{F2} and represents the correction for night-time observations.  We note that  these thresholds correlate very well with the altitude of the station as illustrated in the same figure suggesting that monthly threshold temperatures may be completely determined from the altitude of the stations. From linear fits for each month we established monthly relationships between threshold and altitude of the form: $T_{0}(h) = A\times h + B$ where  $A$ and $B$ are constants for each month with values indicated in Table \ref{T1}. Slight differences along the year are observed for these two parameters with slope or lapse rates variations of the order of $10^{-1}$ $^{\circ}$C and negligible variations in the intercepts. \\

\subsection{Kriging interpolation of altitudes} \label{bozomath}

\begin{figure}
\epsscale{.80}
\plotone{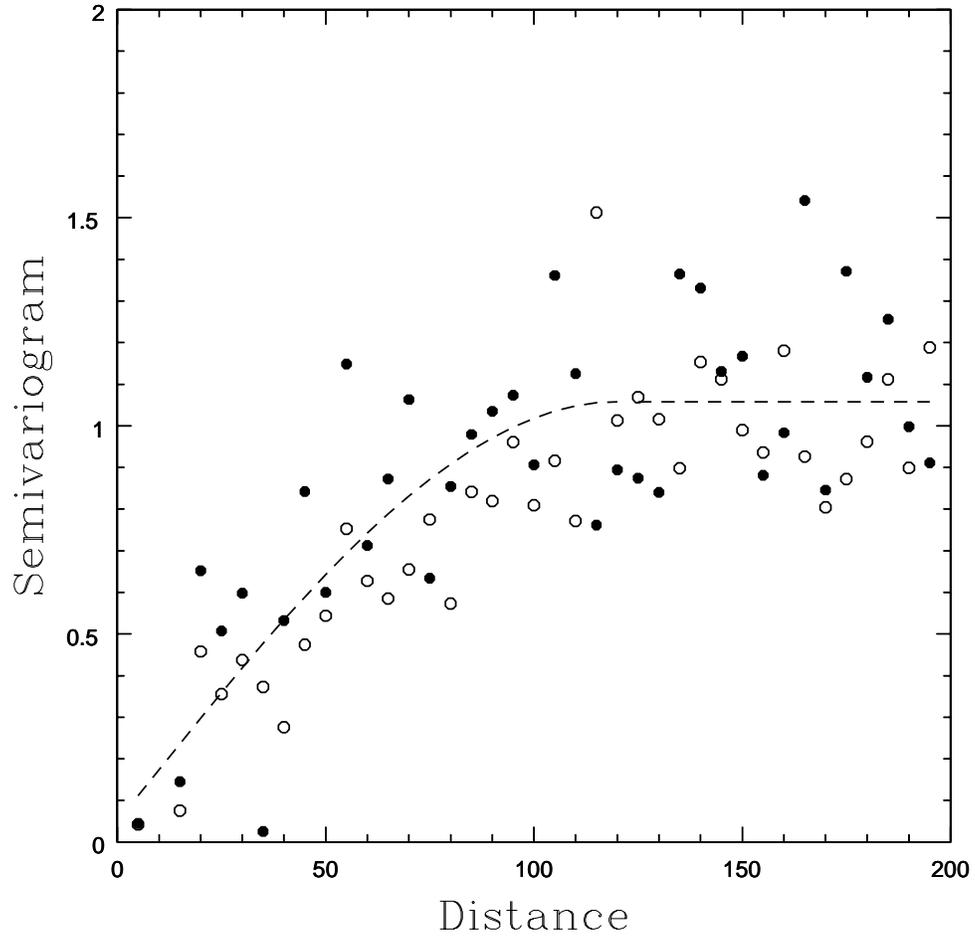}
\caption{Semivariograms computed along south-north (filled circles) and west-east directions (open circles). Both semivariograms are well described by a spherical model (dashed line) with a nugget of $0.048$, sill of $1.01$ and range of $120$.}\label{F3}
\end{figure}

\begin{figure}
\centering
\begin{tabular}{cc}
 {\includegraphics[trim=1.000cm 0.0cm 0.1cm 0.1cm,clip=true,width=0.55\textwidth]{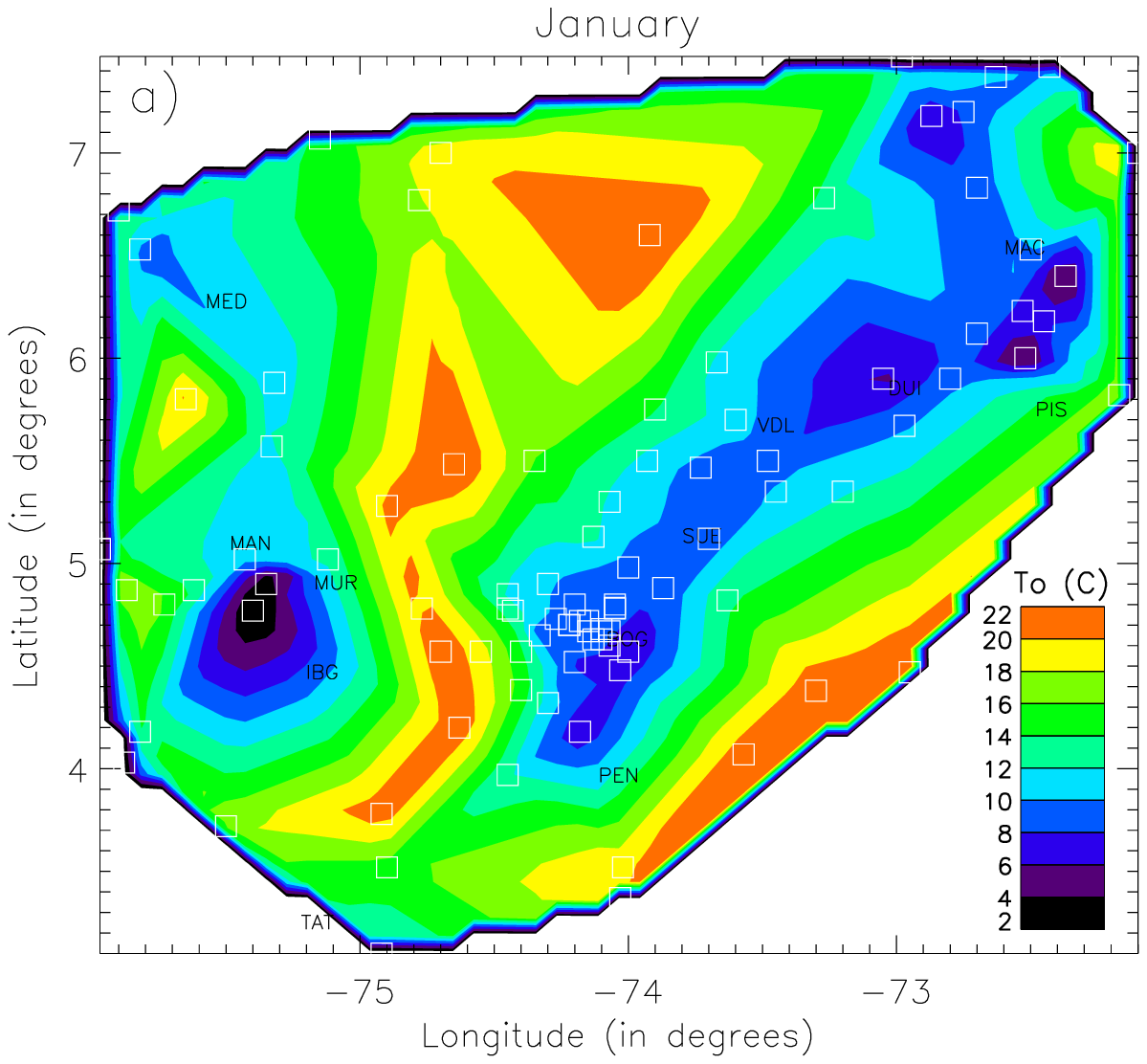}} 
& {  \includegraphics[trim=1.000cm 0.0cm 0.1cm 0.1cm,clip=true,width=0.55\textwidth]{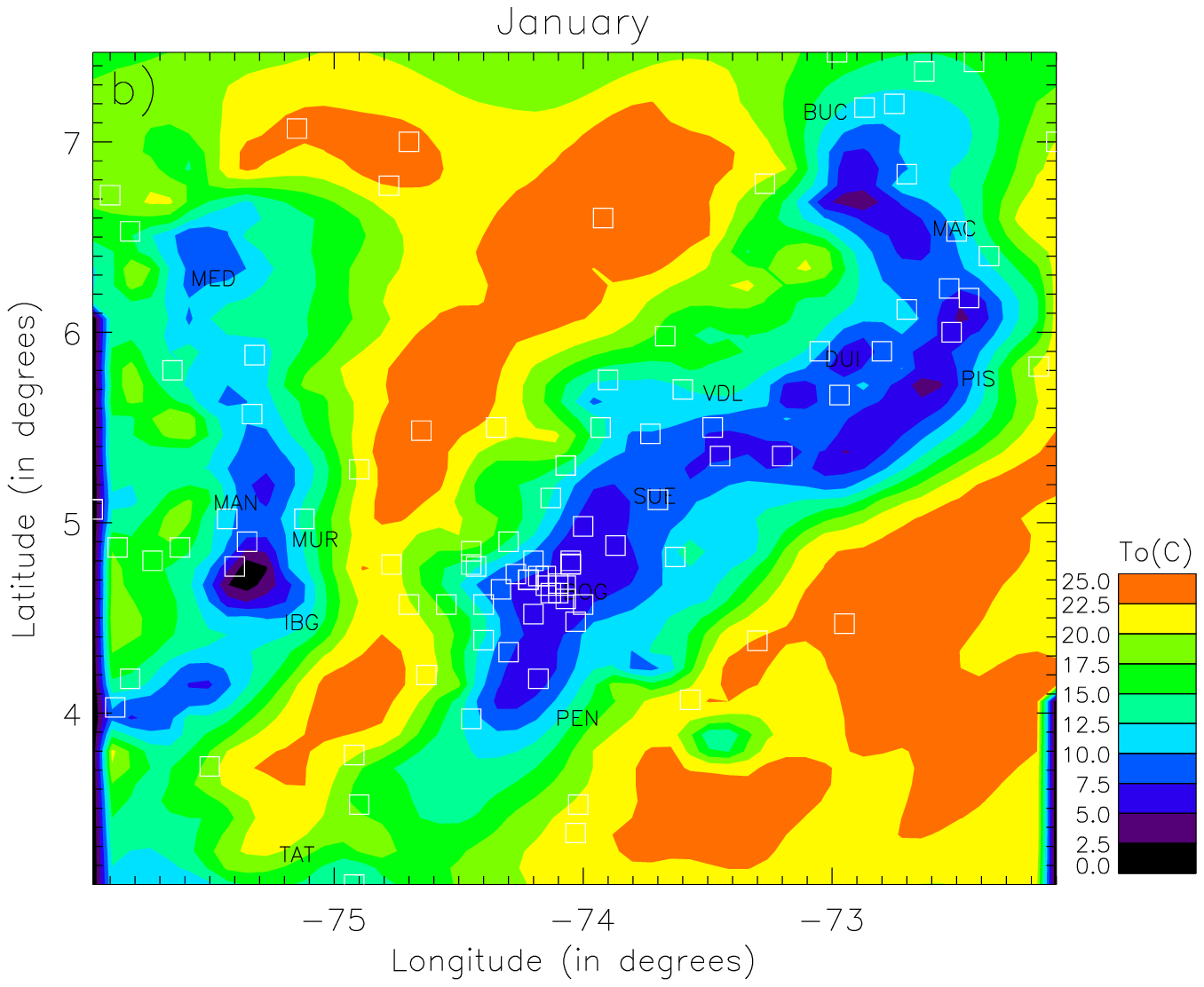}} \\
    {\includegraphics[trim=1.000cm 0.0cm 0.1cm 0.1cm,clip=true,width=0.55\textwidth]{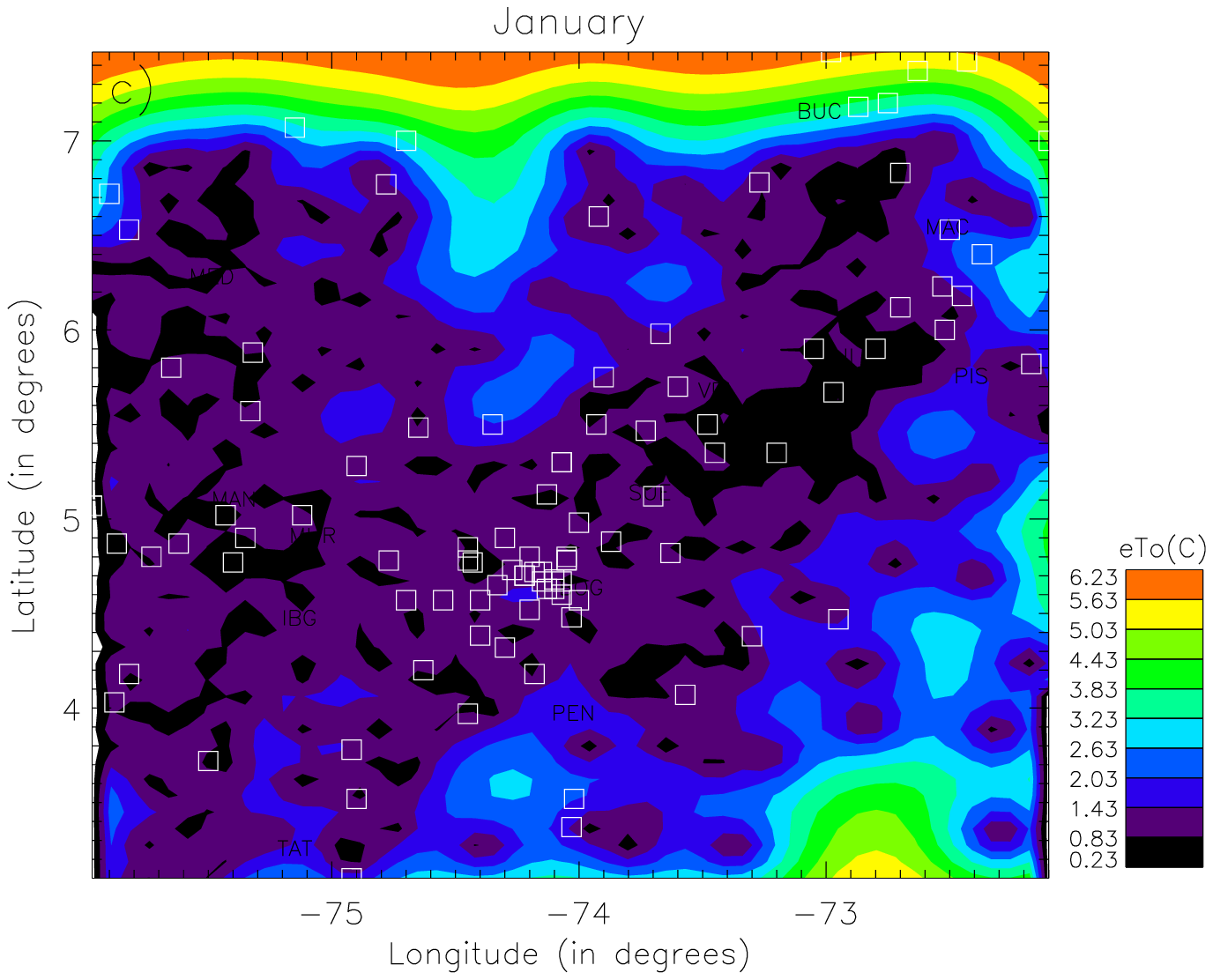} }
 &  {\includegraphics[trim=-2cm -1.5cm -7cm 0cm,clip=true,width=0.55\textwidth]{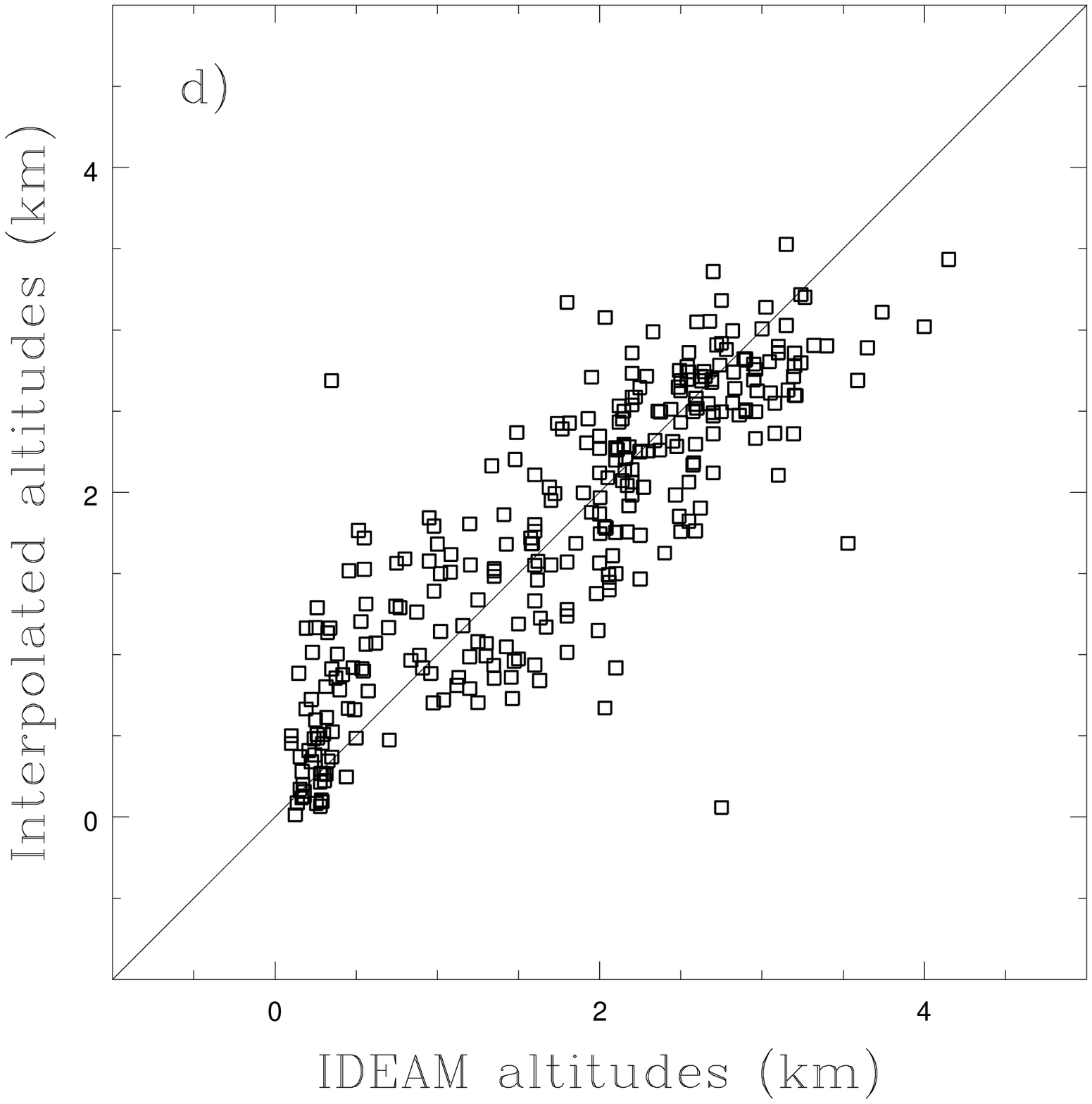}}\\
\end{tabular}
\caption{Monthly threshold temperatures for January in the control region. Panel a) Thresholds obtained through a Delanuay triangulation of the IDEAM long-term records (open squares), b) thresholds using Kriging interpolation of 400 altitudes from IDEAM all-term records, c) sigma errors of the Kriging interpolation and d) cross validation of Kriging  altitudes with IDEAM values. A perfect interpolation is indicated with a solid line.}\label{F4}
\end{figure}

Thresholds in the control region at locations where no weather stations are available require interpolation of the altitudes.  We used the geostatistical method called Kriging, which is a spatial interpolation method that in its simplest form allows to compute altitudes at locations with no data, based on the spatial autocorrelation. This means that two close stations have similar values of altitude whilst two stations farther apart do not. Briefly, the height at the location $s_0$ refered as $\hat{h}(s_0)$ is given by a weighted combination of the IDEAM all records of altitude :

\begin{equation}
\hat{h}(s_0)=\sum_{i=1}^{400} w_i(s_0)\cdot h(s_i)=\lambda_0^T\cdot h
\label{dos}
\end{equation}

where $\lambda_0$ is the vector of Kriging weights $w_i$ and $h$ is the vector of available samples of altitudes. Our sample is defined by the altitudes of all available records of weather stations. The system of equations (\ref{dos}) is equivalent to the  matrix form  $\lambda_0=C^{-1}\cdot c_0$ where $C$ represents the matrix of covariances between all sampled values and $c_0$ the vector of covariances between sampled points and the point in which the value is assessed.  The vector of covariance is obtained from the semivariogram $\gamma (d)$ defined as:

\begin{equation}
\gamma (d) = \frac{1}{2}E[(h(s_i) - h(s_i + d))^2] \label{tres}
\end{equation}

where $d$ represents  the unit distance or lag from which semivariance is determined. The two semivariograms indicated in Figure \ref{F3} were computed for lags up to 200 km along two mutual perpendicular directions. Both semivariograms present similar spatial variances and are well described by a spherical model (dashed line) with a nugget of $0.048$, sill of $1.01$ and range of $120$. The spherical model was found as the most suitable for modeling the covariances (\ref{tres}) and for solving the weights in the system of equations (\ref{dos}) by using GSL-GNU numerical libraries for matrix inversion. A comparison of the interpolated heights with real altitudes from all-term IDEAM weather stations is shown in the Figure \ref{F4}. Cross validation of the Kriging interpolator was done by hiding each IDEAM data point during the computation of the weights and then comparing the prediction with the real value. Differences with an ideal Kriging interpolation (solid line) are within 3$\sigma$ 
which we consider enough good for our purposes.

Based on Kriging results for altitude interpolation in the control region, we computed the monthly thresholds using the values appearing in the Table \ref{T1}. Results are very similar for all the months with small differences due to the bimodal annual cycle. Threshold temperatures for January with resolution of 28$\times$18 km  are shown in the panel b) of Figure \ref{F4}.  The coldest region (0-2)$^{\circ}$C corresponds to the \textit{Nevado del Huila} the highest snow peak in the central branch of the Andes. Thresholds between (2-7)$^{\circ}$C are observed in the eastern Andes comprising \textit{Villa de Leyva} (VDL), \textit{Pisba} (PIS) and \textit{Macaravita} (MAC). Warm regions with thresholds above 20$^{\circ}$C characterize the valley of \textit{Magdalena} river between the central and eastern branches of the Andes and the southestern regions on zero altitude zones. 

For comparison purposes, we show in panel a) the thresholds obtained using Delaunay triangulation of the 63 IDEAM long-term monthly records. We note that the spatial domaine is constrained in this case and defined by the triangulation of the 63 irregularly distributed long-term IDEAM weather stations. Differences between both interpolations are significant mostly in regions surrounded by strong topographic gradients in which a better estimation of the threshold temperatures using Kriging is expected. This is particularly noticed in the poor IDEAM sampled region centered on PIS where a maximum difference of $\pm8^{\circ}$C is observed. Other regions with low number of weather stations are the south of \textit{Bucaramanga} (BUC) and the south-east of \textit{Medell\'in} (MED). On these regions the combination of the lack of weather stations and strong gradients in altitude make the Delanuay interpolation unrealistic. Mean differences of $\pm 3^{\circ}$C over the eastern Andes are atribuited mainly to strong 
gradients as the region is well sampled by the IDEAM data archive. Thresholds obtained from altitude are more precise than those obtained by interpolation of weather station records and are used as a reference for cloud detection in the B4 band.

Uncertainties in the threshold temperature determination are computed using the sigma error of the Kriging interpolation. The largest errors up to 6$^{\circ}$C are observed over the sea level locations and into the valleys while errors over the Andes are always below 1.5$^{\circ}$C as  shown in panel c) of Figure \ref{F4}.  For the control region the thresholds are computed  with an uncertainty of $\pm 2^{\circ}$C.  

\begin{figure}
\centering
\begin{tabular}{cc}
   \includegraphics[trim=3.000cm 0.2cm 3cm 0.5cm,clip=true,width=0.55\textwidth]{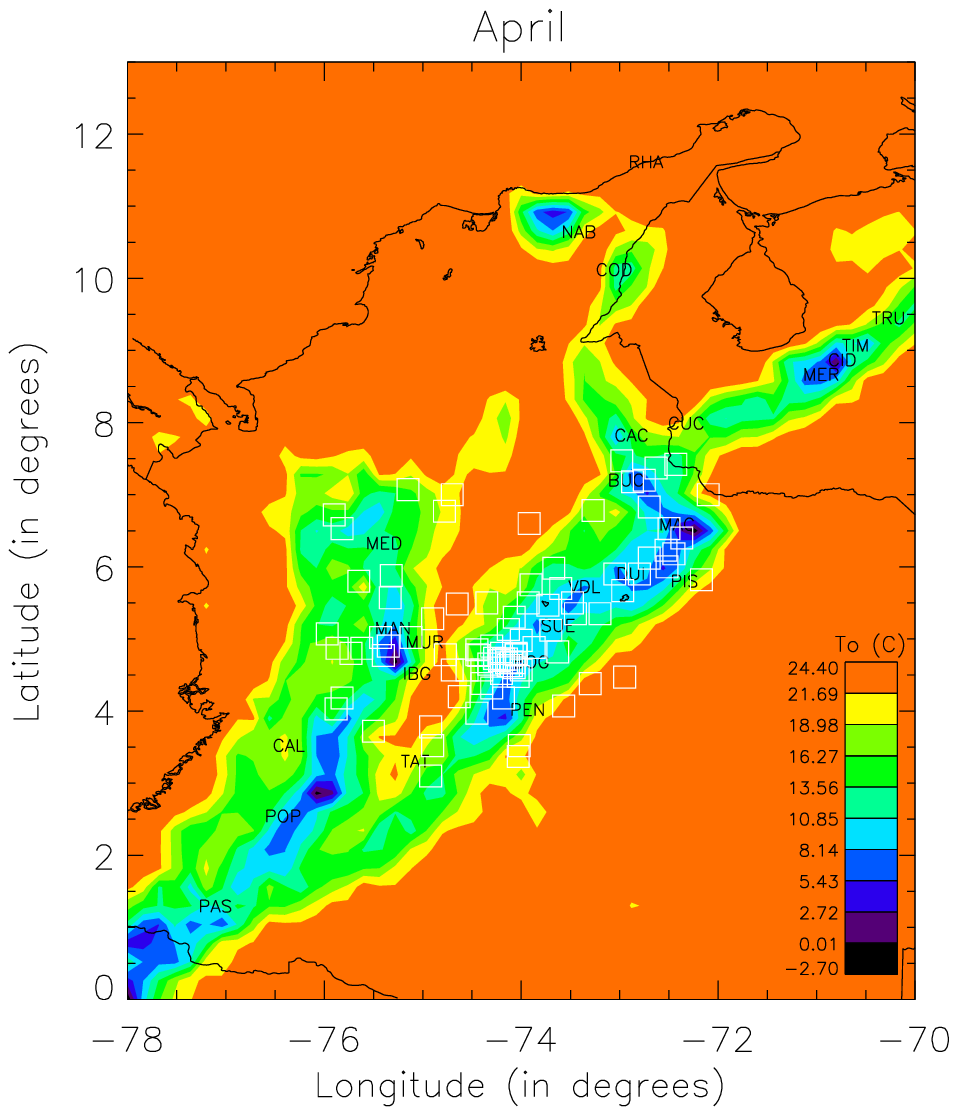} 
   & \includegraphics[trim=3.000cm 0.2cm 3cm 0.5cm,clip=true,width=0.55\textwidth]{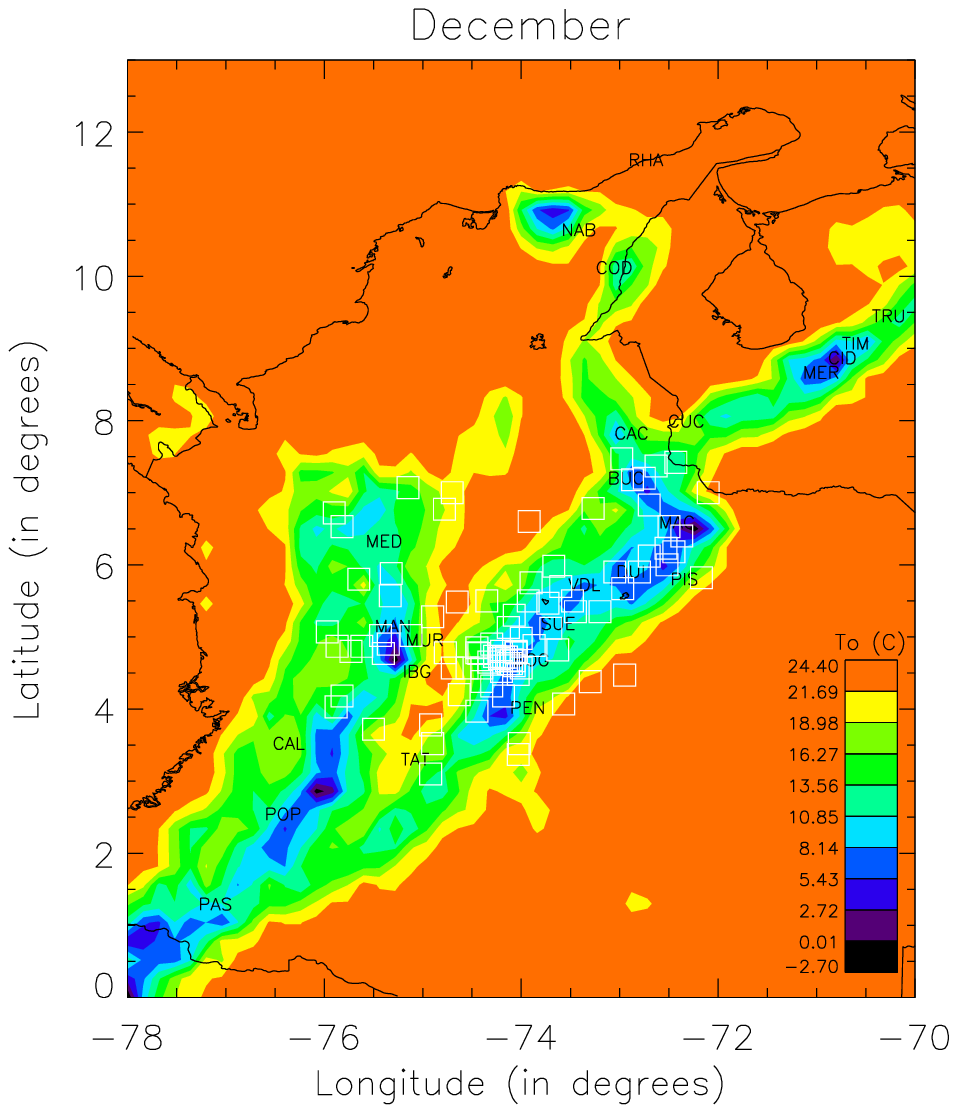}\\
  
\end{tabular}
\caption{Monthly threshold temperatures for cloud detection in the B4 band over the extended region for April and December which are representative of the rainy and dry seasons, respectively. The highest mountains in branches of the Andes shown in blue present thresholds below 5$^{\circ}$C. Caribbean and Pacific coasts, the valleys of the Magdalena river and the flat south-eastern region are marked by thresholds between (20-25)$^{\circ}$C.    }\label{F5}
\end{figure}

With the aim of computing monthly threshold temperatures over a wider region including the \textit{Observatorio Nacional de Llano del Hato} where log-books of cloud cover fraction are available, we apply monthly threshold-altitude relations over what we denoted as extended region covering latitudes between 0 and 13N and longitudes between 70 and 78W. A set of $3000$ altitudes over this region were  obtained from \textit{Google-Maps}  and interpolated using the methodology discussed above. Monthly threshold temperatures between $21$ and $24^{\circ}$C are observed across Caribbean coastal regions and in the valleys of the rivers Magdalena and Cauca in panels of Figure \ref{F5}. In the proximity to the Andes, thresholds get colder specially in the highest mountains indicated with purple in the figure such as the four highests locations over the Colombian and Venezuelan Andes highlands with threshold temperatures below $0^{\circ}$C.  Uncertainties in threshold temperatures over the extended region are assumed to 
be equal to 
those obtained in the control region.

Interpolated threshold temperatures covering the surface of the extended region are then used in our algorithm for cloud detection in the B4 band. For a fixed instant, if the temperature obtained from the GOES image at 10.7 $\mu$m is colder than $T_0$ then the presence of a cloud above the location is inferred.     

\section{Threshold temperatures at the upper troposphere}

Clouds along the upper troposphere are detected following a similar schema, defining first a threshold temperature for high altitudes where the weighting functions of B3 shows the maximum value. We denote the altitude of the atmosphere measured from the ground with $H$, different to the altitude of a weather station $h$. We recall that the weighting functions depending on both, the wavelength and $H$, determine the layer from which the emitted radiation originates. For the B3 band, the median height values of the weighting functions are in the interval $8-10$ km of altitude above the ground with a maximum at 9 km (Cavazzani et al.2011). 

Threshold temperatures at upper troposphere may be obtained using different approaches. Hydayat et al. (2012) considered the reference atmospheric model of Kneizys et al. (1996) which provides a vertical tropospheric profile of temperature for latitudes within $\pm15^{\circ}$ and adopted  a threshold equal to $-30.15^{\circ}$C at the observatory location based on a $\chi^2$ analysis with data of sky visual inspection. Our approach is based on the use of three thresholds each one at heights of $8$, $9$ and $10$ km obtained from radiosonde upper-air measurements.

We make a strong emphasis on the reliability of the temperature at upper troposphere. Particularly, we are interested in finding trustworthy threshold values between 8 and 10 km. Data were obtained twice a day from the freely available database of the University of Wyoming\footnote{http://weather.uwyo.edu/upperair/sounding.html} which provides information of temperature, humidity, pressure and mixing ratio as a function of the altitude for two stations located inside the extended region, \textit{80222 SKBO} (74$^{\circ}$9'0"5W, 4$^{\circ}$42'0"N, 2546m) in central Andes and \textit{78988 TNCC} (68$^{\circ}$57'36"W, 12$^{\circ}$11'60", 54m) in the Caribbean coast. Upper-air temperatures of both stations exhibit similar patterns as a function of the altitude. Around 60 radiosonde records were used per month adding data from both stations. Altitude resolution is variable with typical values ranging from 200 up to 600 m.

\begin{center}
\begin{figure}
\begin{center}
\begin{tabular}{cc}
{\includegraphics[width=8.5cm,height=8.0cm]{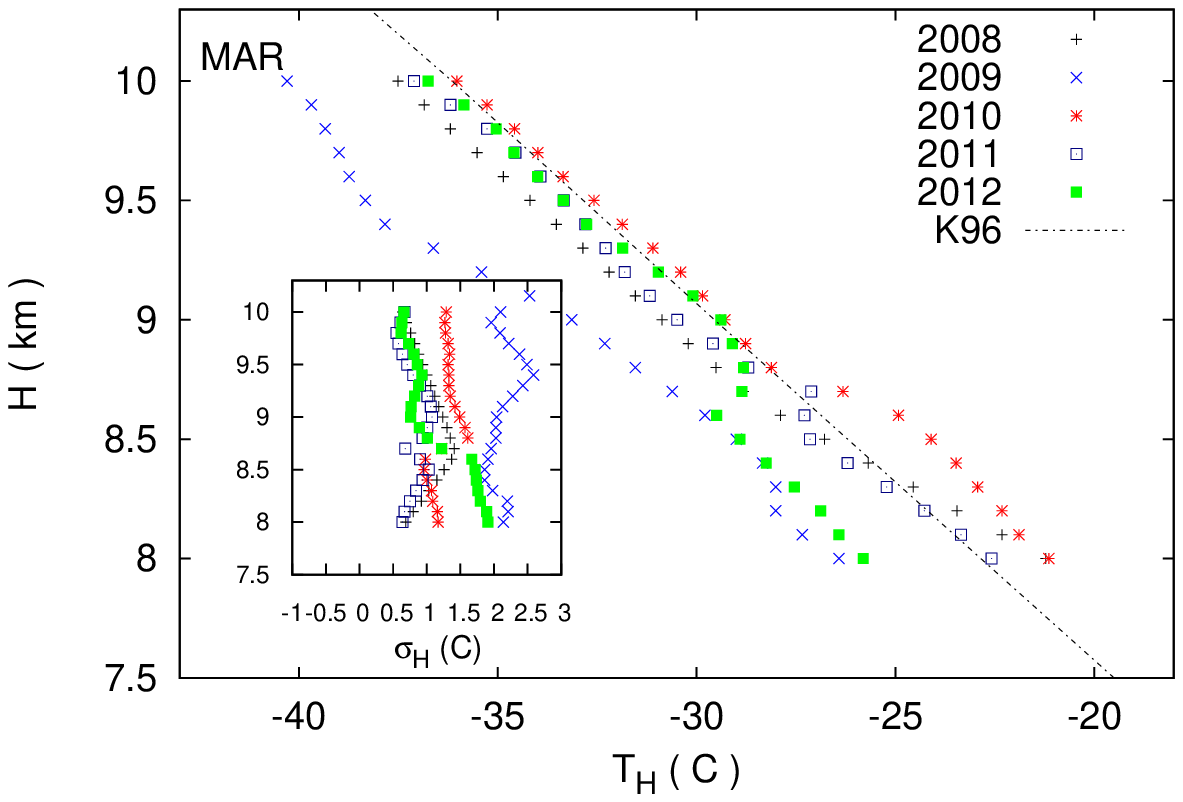}}   
& {\includegraphics[width=8.5cm,height=8.0cm]{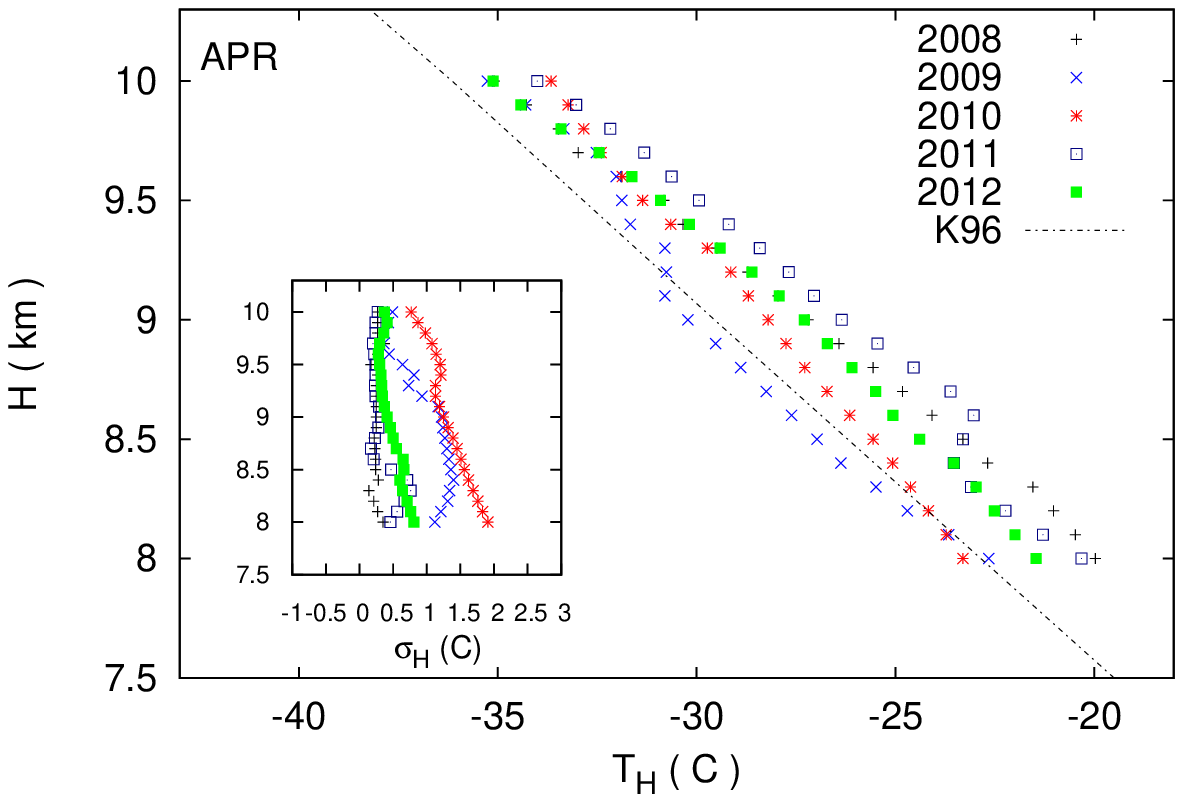}} \\
 {\includegraphics[width=8.5cm,height=8.0cm]{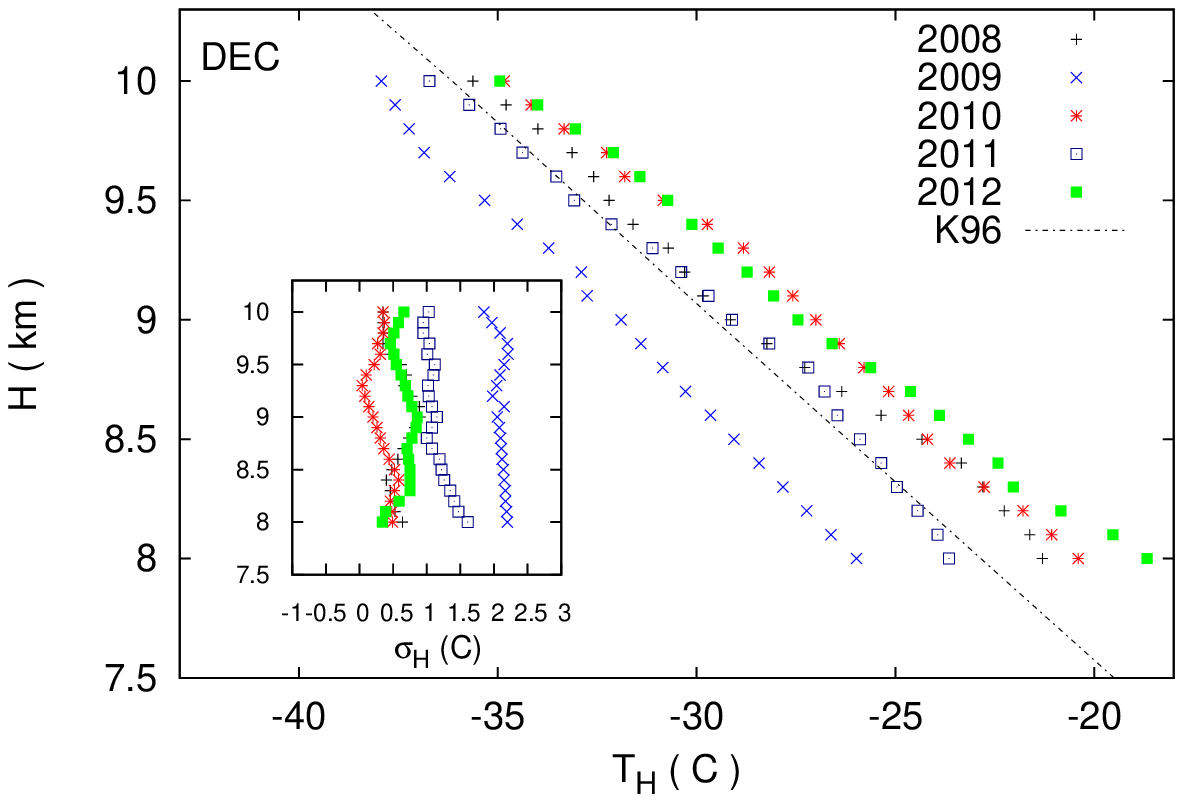}} 
 &   {\includegraphics[width=8.5cm,height=8.0cm]{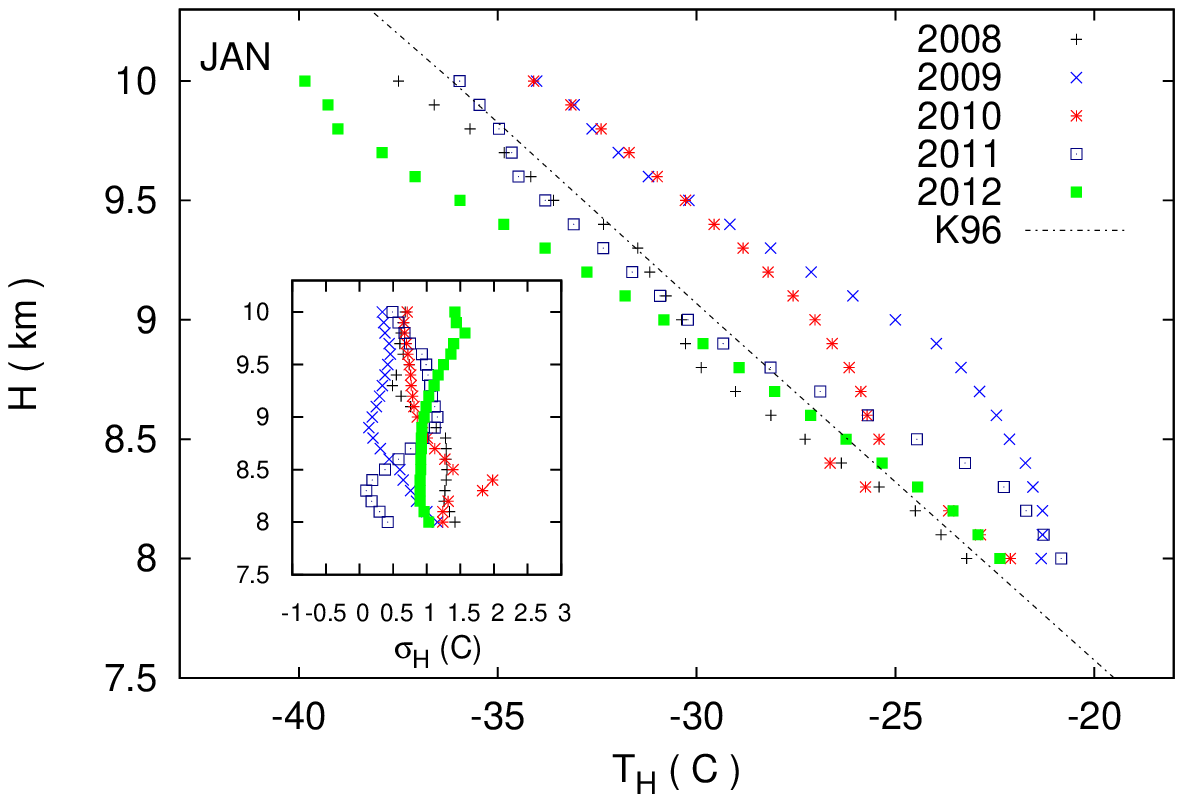}}\\
 \end{tabular}
\caption{Monthly threshold temperatures for cloud detection in the B3 band over the extended region for March and April which are representative of the rainy season and for and December and January representative of the dry season. Yearly standard deviations in the upper atmosphere obtained from radiosonde data are also displayed in the interior plot. Temperature from the tropical atmospheric model of Kneizys et al.(1996) is indicated with a solid line.}\label{F6}
\end{center}
\end{figure}
\end{center}

We compute monthly mean temperatures by considering measurements made twice daily (00 and 12 UTC) during the years  2008-12 at each $H$  using linear interpolation along a uniform vertical grid of 21 points between $8$ and $10$ km. In order to follow the same approach than the one used for ground thresholds, we define the threshold temperature at the altitude $H$ as $T_{H} \equiv <T_{H}> -  3\sigma_{H}$ where $<T_H>$ is the monthly mean value and $\sigma_{H}$ is the standard deviation. Monthly thresholds as a function of the altitude are displayed in Figure  \ref{F6}. Monthly standard deviations change with altitude and time showing yearly variations up to $\pm 2^{\circ}$C in the worst of cases. In the tropics, the temperature anomalies at high altitudes is enhanced by the presence of moderate size clouds usually associated with shallow convection responsible for anomalies up to  $+2^{\circ}$C at altitudes from 5 up to 14km (Folkins 2008). This tropospheric warming has been largely observed and we suggest 
that is the main source of the sigma variations.

\begin{figure}
\centering
\begin{tabular}{ccc}
   \includegraphics[trim=3.00cm 0.05cm 3cm 0.4cm,clip=true,width=0.34\textwidth]{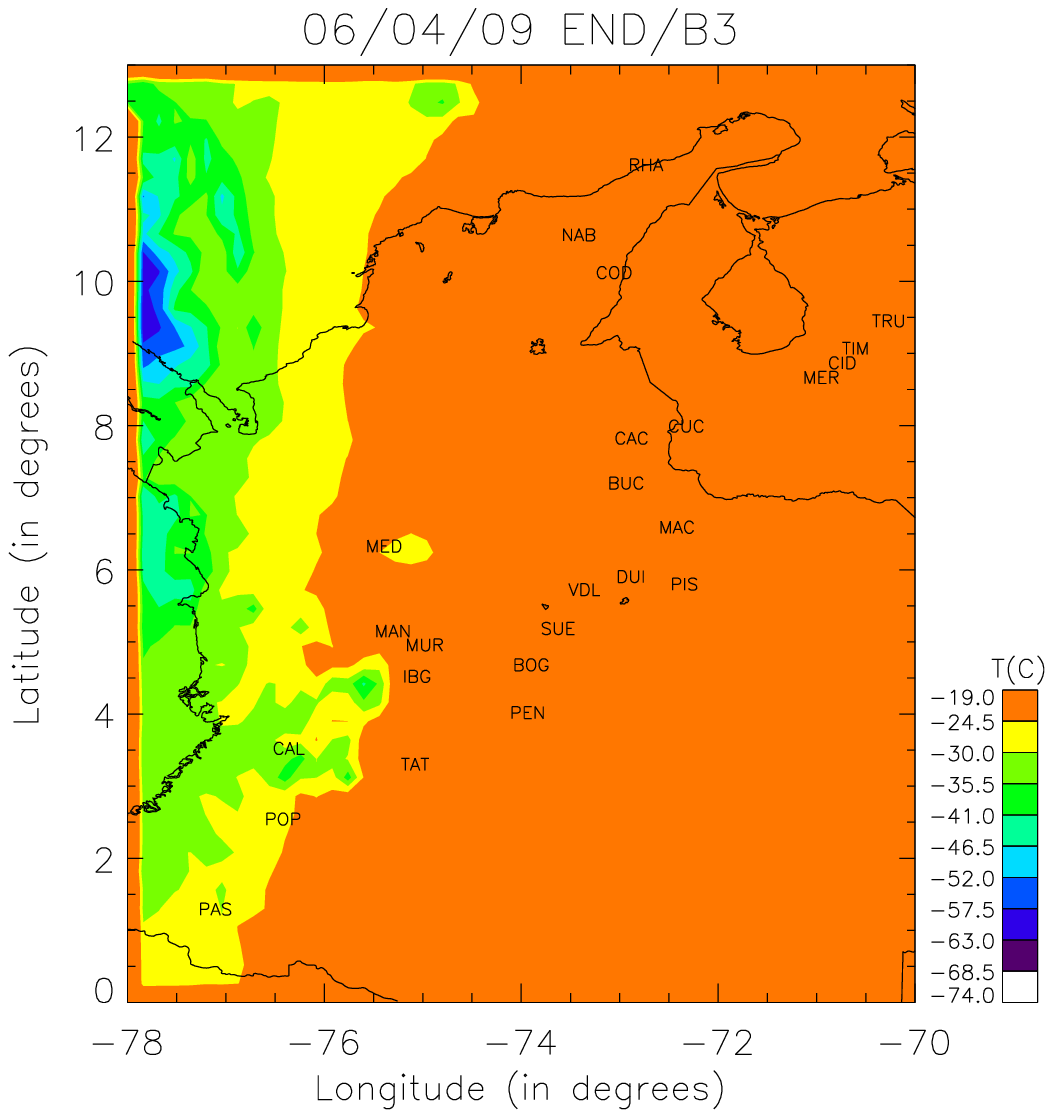} 
   & \includegraphics[trim=3.00cm 0.05cm 3cm 0.4cm,clip=true,width=0.34\textwidth]{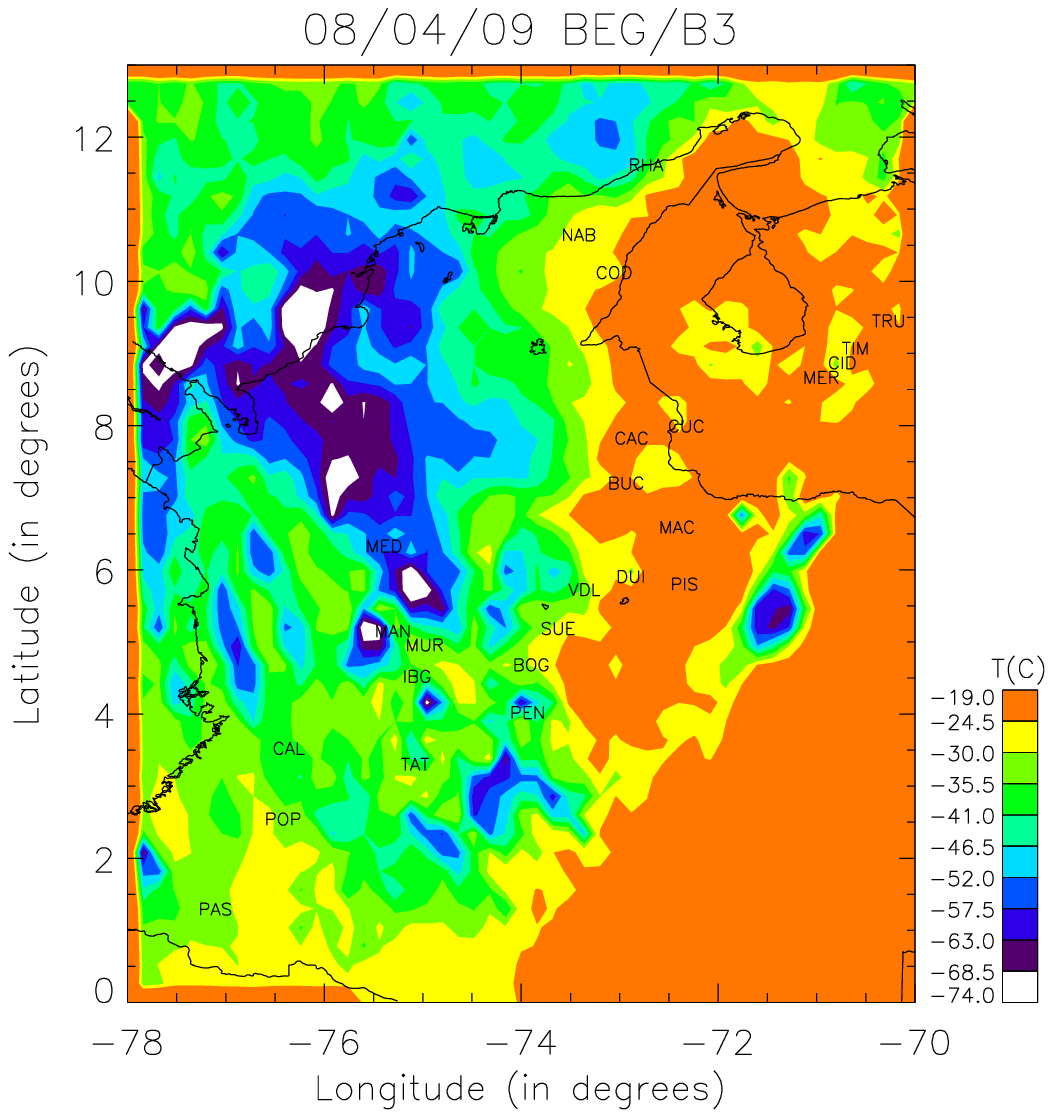}
   & \includegraphics[trim=3.00cm 0.05cm 3cm 0.4cm,clip=true,width=0.34\textwidth]{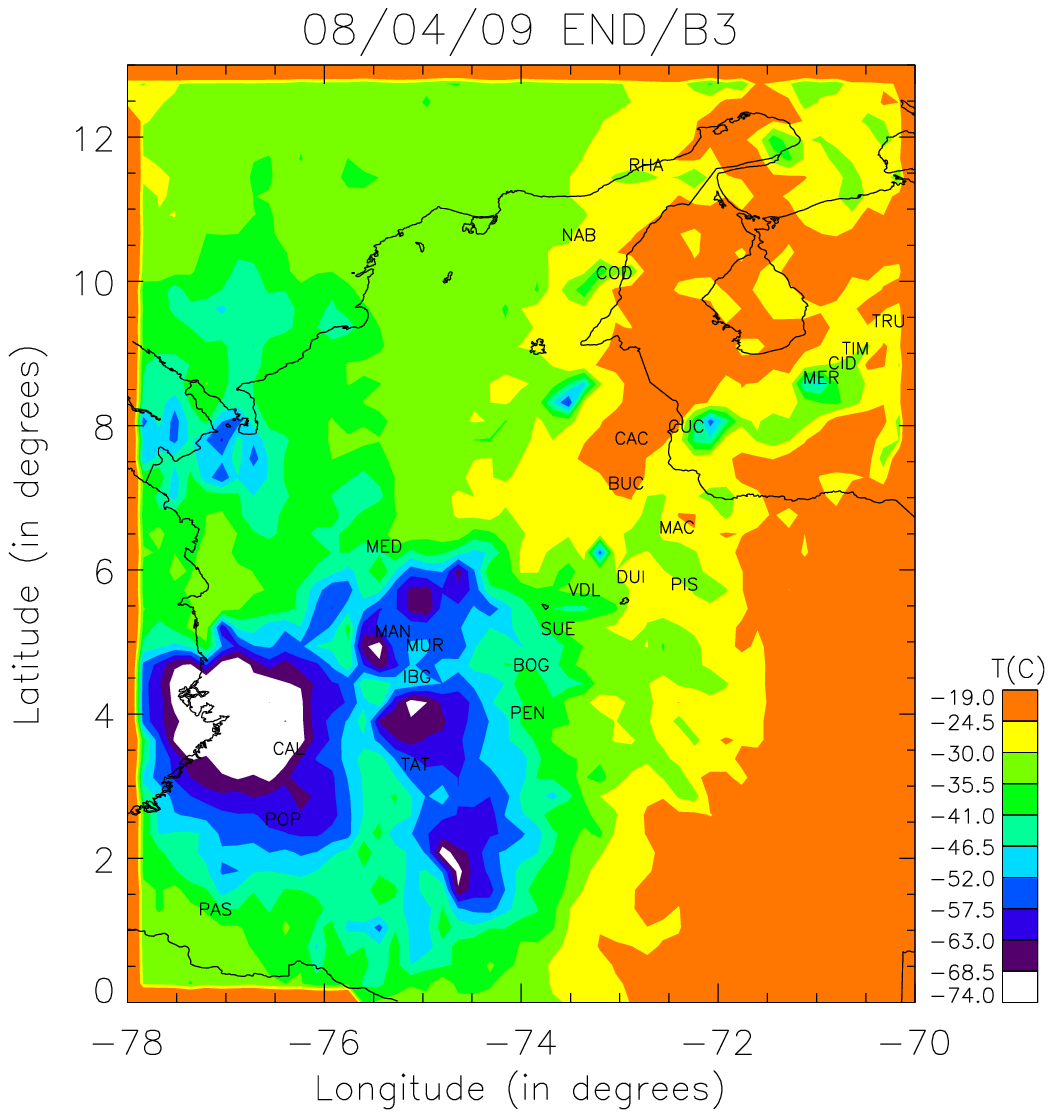} \\
   
\end{tabular}
\caption{Temperature at upper atmosphere using B3 radiances. High altitude clouds in blue move from left to right in timescales of days before being blocked by the Andes system, leaving clear regions mostly in the western Venezuela and in the eastern Andes of Colombia.}
\label{F7}
\end{figure}

The further assumption that these thresholds describe the upper troposphere of the whole extended region in many aspects is a strong hypothesis, however the area of study is affected by atmospheric phenomena including the westerly low level \textit{Choc\'o} jet coming from the Pacific coasts of Colombia (Poveda \& Mesa 2000) and the Caribbean trade winds which leads to similar daily patterns of high altitude clouds. Middle and right panels in Figure  \ref{F7} show a typical evolution during the night of 8th of April 2009. We see two images in B3 taken at the beginning and at the end of the night. High altitude clouds shown in green color in the left panel coming from the Pacific ocean are blocked by the Andes system leaving clear zones in the Andes of Venezuela and over the Amazon zones.

We note that the fact of extrapolating thresholds over all the extended region, the possible changes in the instrumentation and the differences between night and daytime observations may affect the observed temperatures and therefore introduce considerable noise in the computed thresholds.  However, the monthly thresholds obtained with these two stations are in agreement with the predictions of the tropical atmospheric model of Kneizys (1996) indicated with the solid line in Figure \ref{F6}.

\section{Methodology and night classification }

We implemented an algorithm for percentage of clear sky calculation during the night through a comparison pixel to pixel of GOES 12/13 interpolated temperatures with monthly threshold values obtained from long-term records of weather stations and radiosonde databases. We constructed four monthly threshold images, one for band B4 labeled as $T_0$ and three for B3 labeled as $T_H$ each one at heights $H=8,9,$ and $10$ km. Daily comparisons were done in both bands at 00:15:15, 06:45:13 and 09:45:14 UTC with similar approach to those described in Hidayat et al. (2012) and  Erasmus \& Sarazin (2002). First we used daily radiances in the B4 band to compute brightness and actual temperatures. If the satellite derived temperature is colder than the monthly threshold at B4 the presence of middle high clouds above the pixel image at the instant $i$ is inferred :

\begin{eqnarray}
\mbox{if } T_{B4,i}  \geq    T_{0}      \longrightarrow  clear \nonumber \\
otherwise \longrightarrow  covered     
\label{critb4}
\end{eqnarray}

where $T_0$ is a monthly constant depending only on the altitude of the terrain. If a pixel is classified as covered in B4 it will be also covered in the B3 band. On the contrary, if the pixel is clear in B4 we still have to verify the presence of high altitude cirrus clouds by doing  a further and similar analysis using the B3 band. The pixel is classified as clear or covered in B3 at the altitude $H$ in the following way : 

\begin{eqnarray}
\mbox{if } <T_H>   - T_{B3,i}   \leq   3 \sigma_H    \longrightarrow  clear \nonumber   \\
otherwise \longrightarrow  covered 
\label{critb3}
\end{eqnarray}

where $<T_H>$ denotes the monthly mean temperature of the upper troposphere at height $H$ and  $\sigma_H$  the standard deviation at that height. We compute three values of clear sky fraction at the instant $i$ and at each altitude $H$ using inequalities (4) and (5) by counting the number of clears or covered pixels.  The total clear sky fraction at altitude $H$ is then obtained from the average of the three instants $i$ along the night. One value of CSF is computed in this way at each altitude.

\subsection{Classification of the nights}

The nights in the extended region are classified as spectroscopic or photometric at each altitude $H$ based on the values of total clear sky fraction.  We define the quality of the night, at the height $H$, as spectroscopic if the total clear sky fraction  is larger than $33\%$. As we computed the total CSF with three values, spectroscopic nights are those with $1/3$ of the night free of clouds at least. Sites that fulfill this requirement are more suitable for spectroscopic observations than for photometric ones as long as spectra can be acquired quickly in comparison with photometeric data. We note that this definition is based only on the amount of clear sky and does not involve any sky quality indicator. Spectroscopic Night Simulations (hereafter, SNS) provide monthly averages of clear sky fraction at three different altitudes for each pixel of the image satisfying the condition that CSF is larger than 33\%.

Nights with photometric quality on the other hand, are those with total sky fraction larger than 67\% corresponding to $2/3$ of clear sky during the night at least. If this condition is fulfilled in a particular site, we argue than at the altitude $H$ will be possible to conduct photometric observations. Photometric Night Simulations (hereafter, PNS) provide monthly maps of clear sky fraction at three different altitudes for each pixel of the image with the condition that CSF is larger than 67\%. 

\section{ Validation of the algorithm}

Based on total CSF values obtained with SNS and PNS outputs, monthly averages at altitude $H$ were computed. Cloud cover along the upper-troposphere is sensitive to the altitude, being larger at low altitudes mostly at $H=8$ km, indicating that high cirrus does not affect considerably the net cloud fraction. We used the monthly CSF averages at $H=8$, $9$ and $1$0 km to compute a weigthed mean for the three altitudes at each instant during the night.  Weights were computed through a comparison of the amount of clear sky fraction for photometric nights coming from our model with those obtained from daily number of observing hours reported by log-books of CID or Site 2 in Venezuela. Site 2 is placed at 3600 meters height and enclosed by higher mountains as \textit{Pico Bol\'ivar} (4978m) at less than 40 km away. As the resolution is 28 km in longitude and 18 km in latitude, the threshold temperature at Site 2 is highly affected by the altitude of the closest higher mountains giving colder thresholds than 
expected. Using PNS instead of SNS makes more precise validation because PNS output is  closer to the condition of total clear night. 

\begin{center}
\begin{figure}
\begin{center}
\begin{tabular}{ccc}
{\includegraphics[width=5.0000cm,height=5.0000cm]{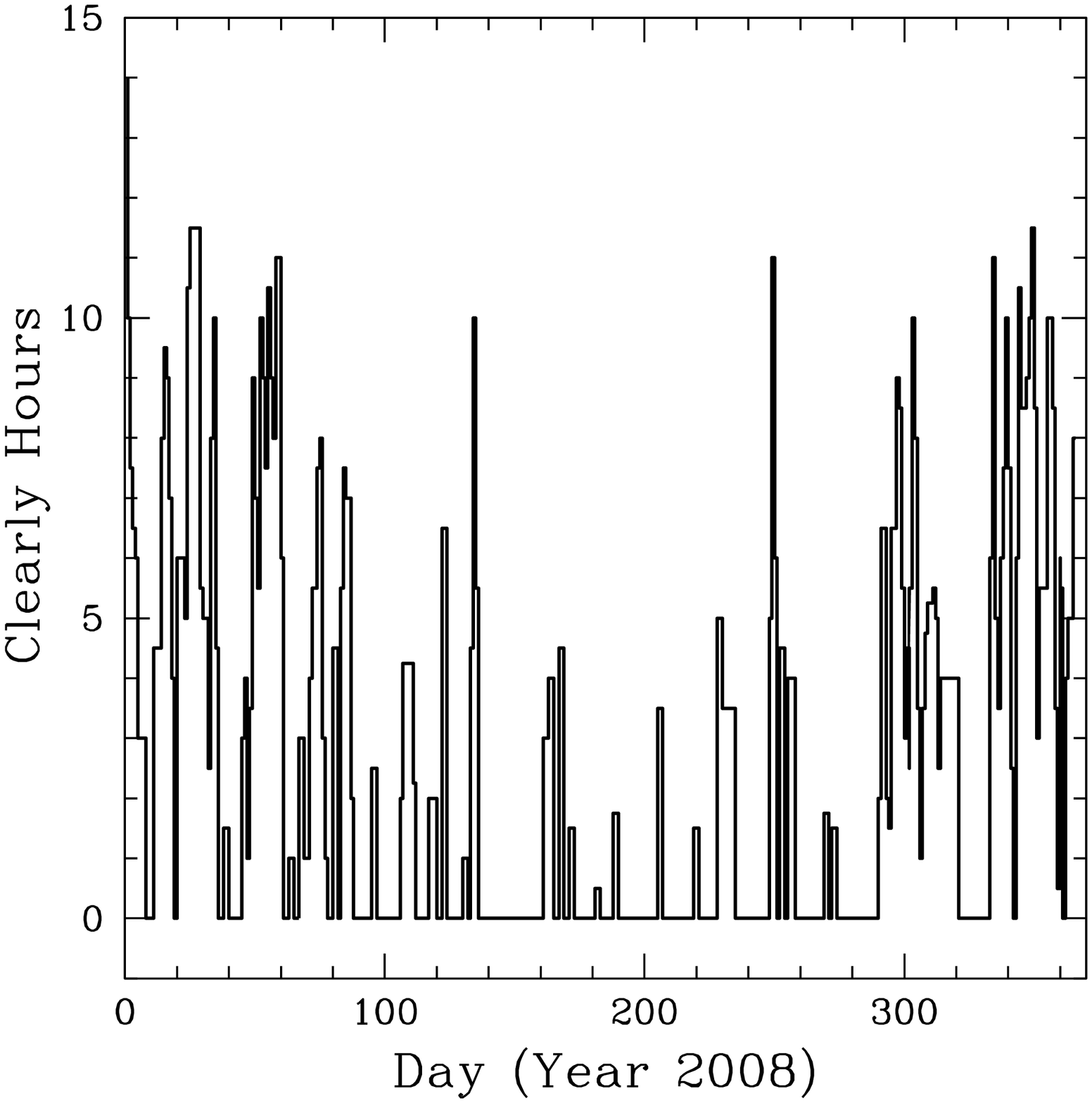}} &
 {\includegraphics[width=5.0000cm,height=5.0000cm]{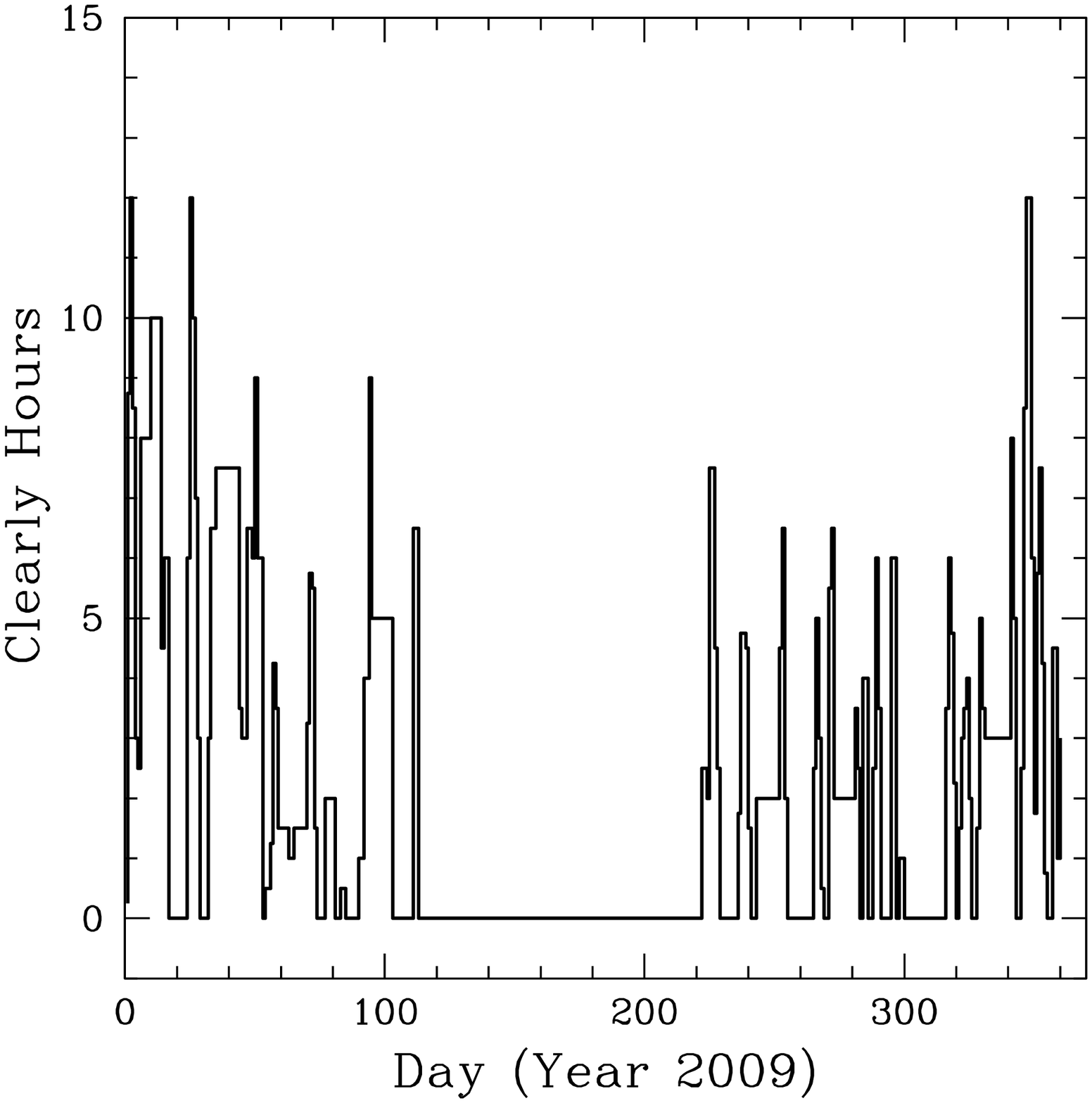}} &
{\includegraphics[width=5.0000cm,height=5.0000cm]{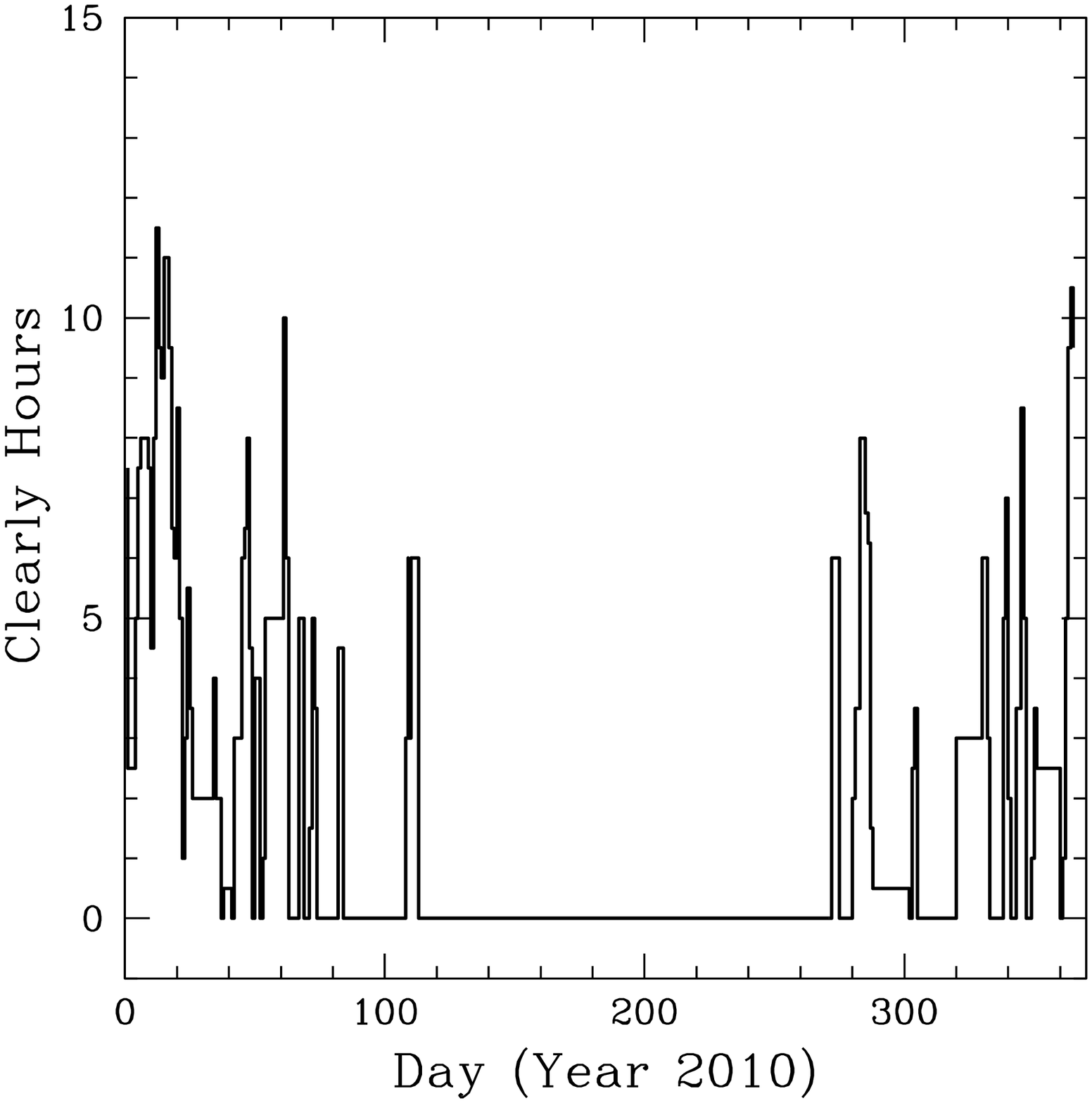}} \\
 {\includegraphics[width=5.0000cm,height=5.0000cm]{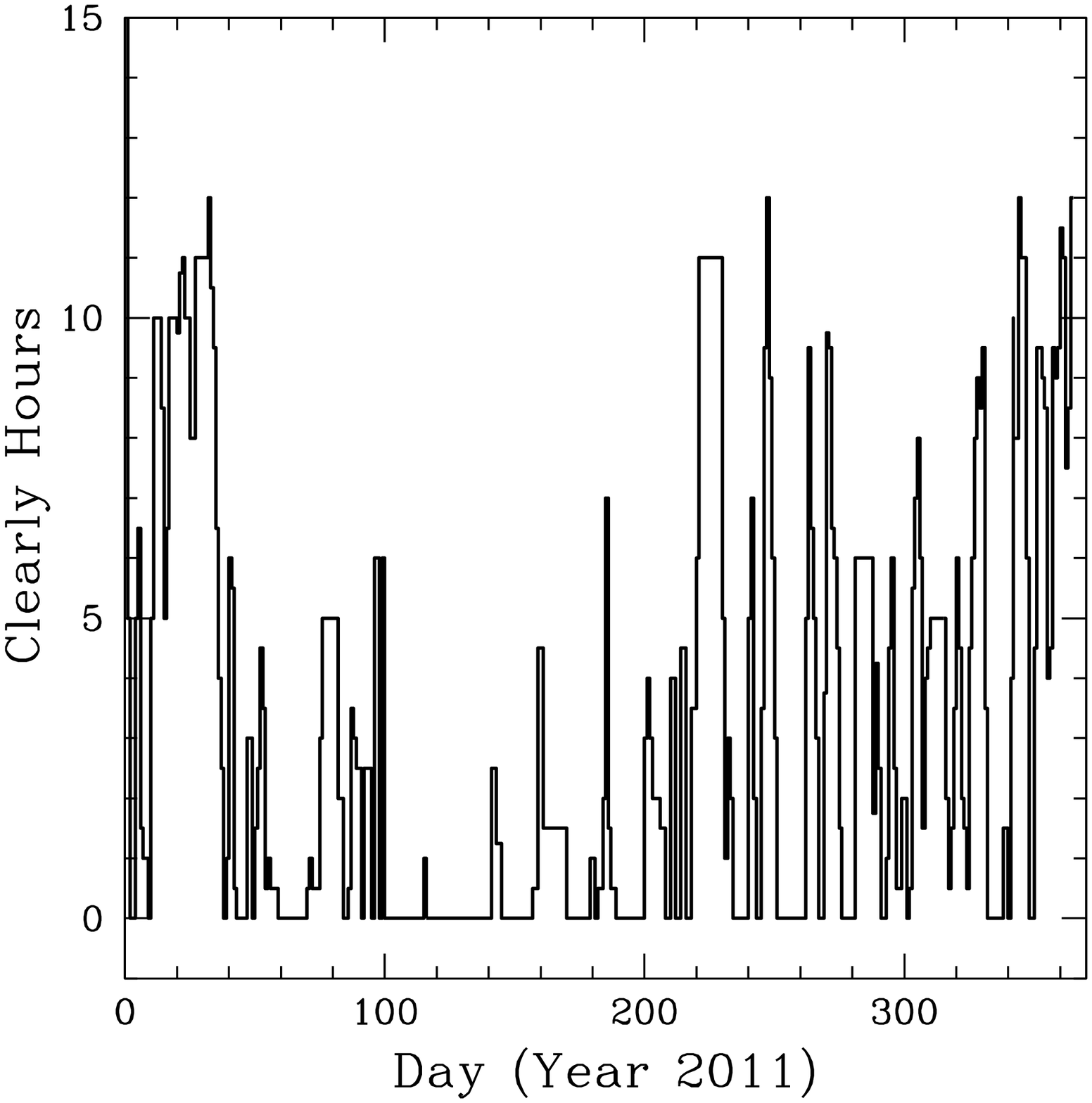}} &
{\includegraphics[width=5.0000cm,height=5.0000cm]{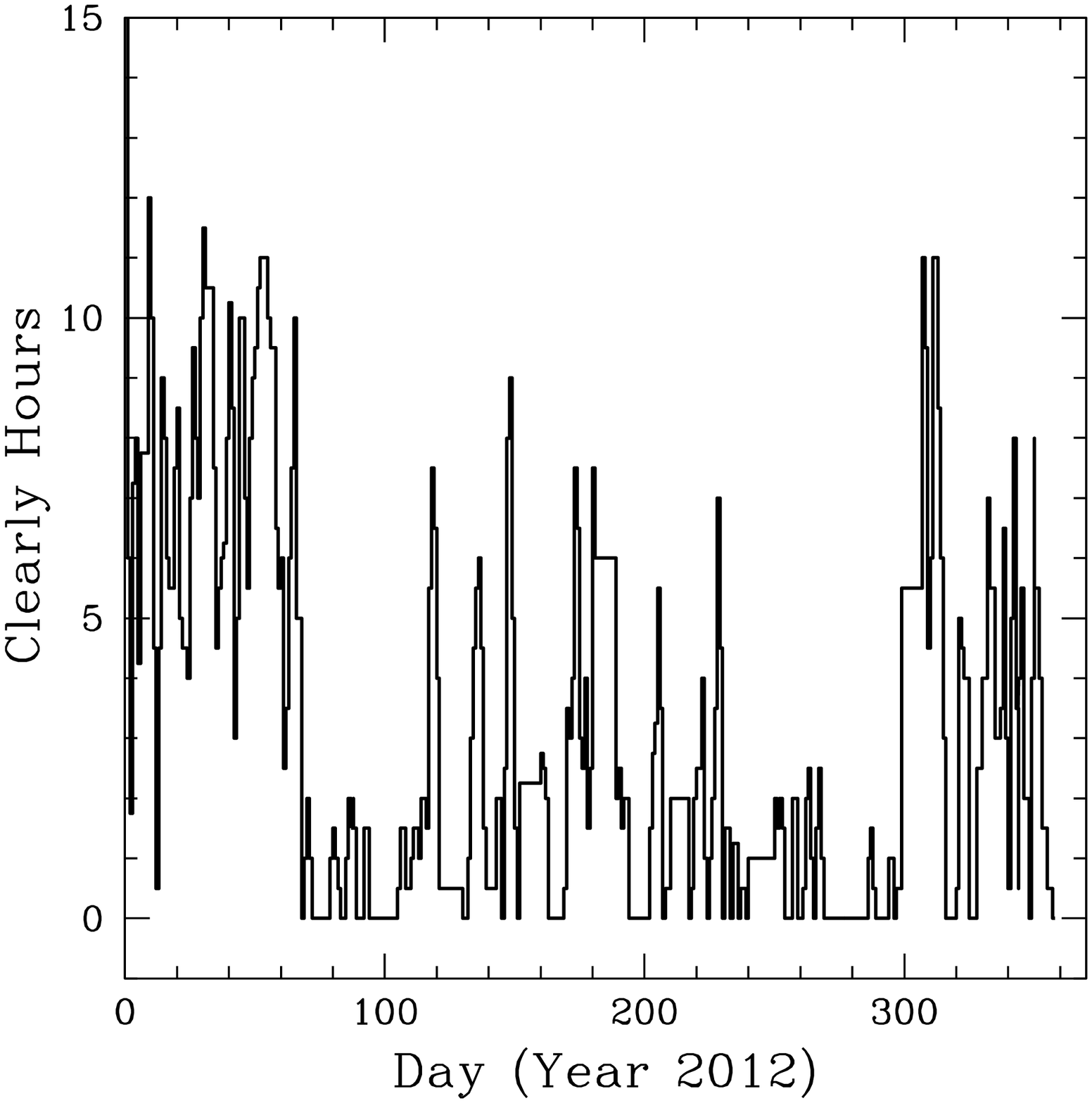}} & \\
\end{tabular}
\caption{Number of daily observing hours obtained from the log-books of \textit{Observatorio Nacional de Llano del Hato} (CID or Site 2) for the years 2008-12. }\label{F8}
\end{center}
\end{figure}
\end{center}

Figure \ref{F8} corresponds to the clear hours above the reference Site 2 or CID as reported in the log-book. Despite the bias present in the log-books  and the missing data for the years 2009-10,  due mainly to bad weather, we did not find considerable differences between observed and expected monthly clear sky fraction values when different weights were used. Therefore, for the purposes of this study, we consider sufficient to use equal weights for the three altitudes.

\begin{center}
\begin{figure}
\begin{center}
{\includegraphics[width=11cm,height=15cm,angle=-90]{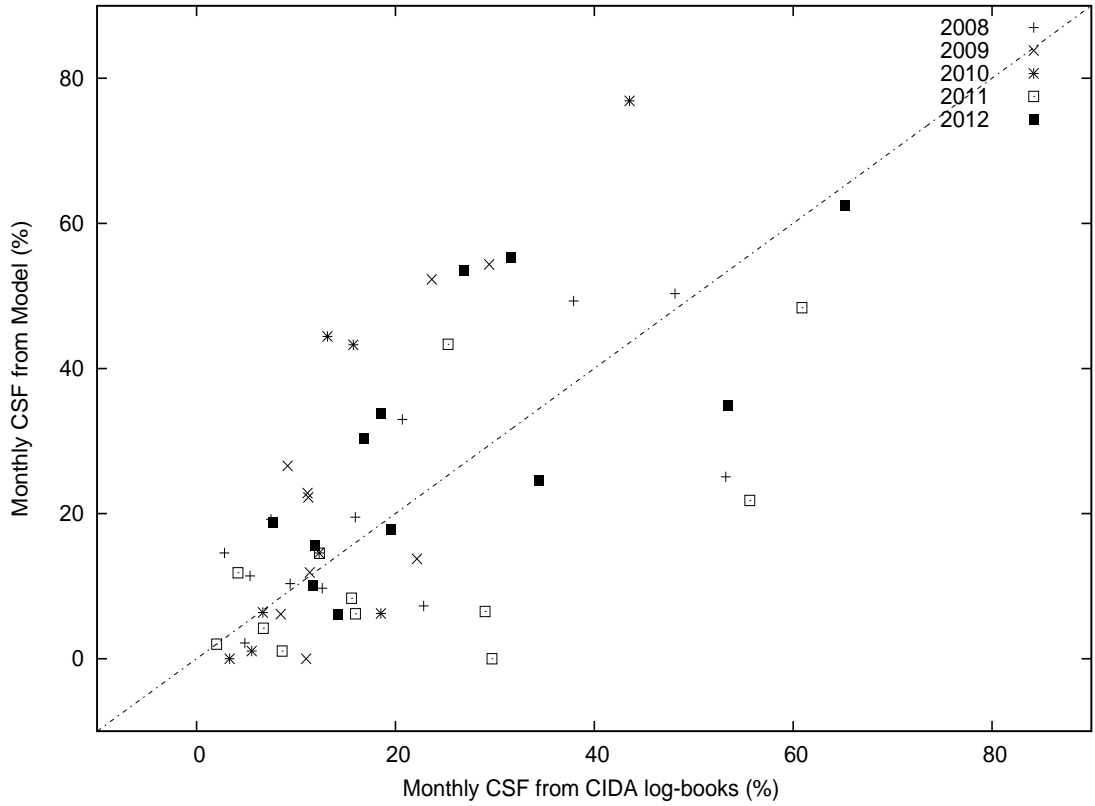}} \\
\caption{Absolute difference between clear sky fraction from our model and those obtained from long-term records of the number of observing hours for 2008-12 at CID or Site 2. A linear regression leads to a slope of (0.76$\pm$0.13) corresponding to a percentage difference of 24\% in comparison with the perfect agreement between  observed and expected CSF values.}\label{F9}
\end{center}
\end{figure}
\end{center}

The validation of our code is limited due to factors such as the lack of completeness of daily log-books, tecnhical issues, high humidity, strong winds  and uncertainties of our model. We used all the information available in the log books as the general description of the night done by the astronomer to compute the amount of daily clear sky fraction for each month of the years 2008-12 at Site 2. Comparison with our monthly results using PNS is displayed in Figure \ref{F9}. The solid line represents no differences between predicted and measured monthly clear sky fraction values. There is a large spread during the considered years, especially for clear sky fraction values between 20-60\%, but in general terms they are in reasonable agreement.

\section{Monthly clear sky fraction for 2008-12}

Monthly percentages of clear sky for the years 2008-12 over the extended region were obtained using the method described above. The main common feature is that monthly averages display maxima at the same locations along the five years. Maximum monthly percentages of clear sky are observed during two periods in the years 2008-12, one from December to February and another from June to July according with the bimodal annual cycle of precipitation (Poveda et al. 2011). The  year 2012 stands out as the best year displaying the highest percentages during both periods.

\begin{figure}
\centering
\begin{tabular}{cc}

    \includegraphics[trim=3.000cm 0.2cm 3cm 0.5cm,clip=true,width=0.55\textwidth]{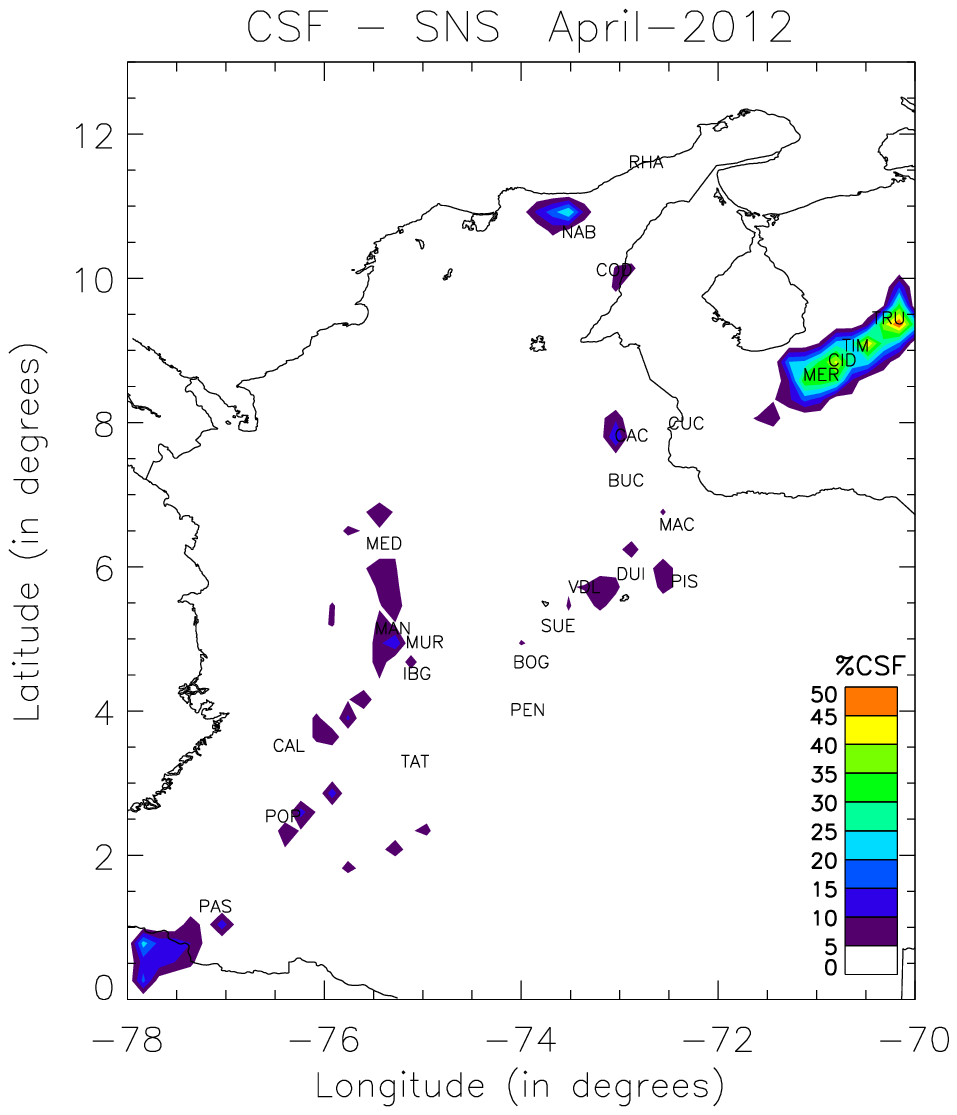}
     & \includegraphics[trim=3.000cm 0.2cm 3cm 0.5cm,clip=true,width=0.55\textwidth]{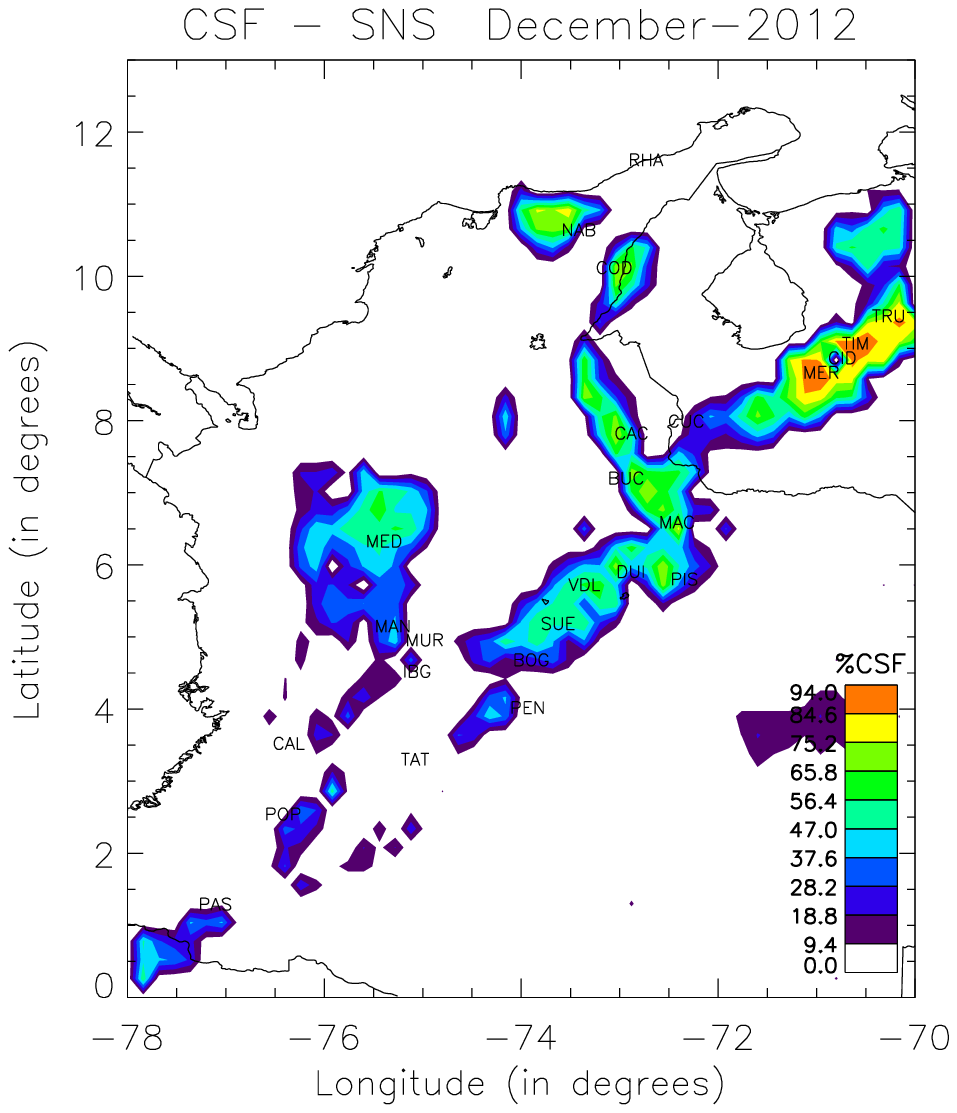} \\
   
\end{tabular}
\caption{Monthly clear sky fraction percentage for spectroscopic nights (SNS) in the extended region for the year 2012 for April and December which are representative of the rainy and dry seasons, respectively. }\label{F10}
\end{figure}

We take the year 2012 as representative of the monthly distributions of clear sky. As illustrated in Figure \ref{F10} our monthly averages of CSF for SNS nights exhibit differences along the year with the highest CSF values always located over a wide region in the central Venezuelan Andes with a maximum CSF$\geq$80\% at Site 1 during both periods. For Site 2 or CID corresponding to \textit{Observatorio Nacional de Llano del Hato} we get a CSF mean value of 58.5\%,  lower in comparison with Site 1 during both periods. 

We have found two regions in Colombia with  similar night qualities to those observed at Site 2. The northern part of the \textit{Sierra Nevada de Santamarta} close to \textit{Nabusimake} (NAB) and the central \textit{Serran\'ia del Perij\'a} toward the east of \textit{Codazzi} (COD). Percentages at these two regions are in the range of 80-90\%  during the first period, 50-70\% during the second and below 30\% from March to May and between August and October. Sites over this region contain the best places in Colombia found in this study. 

Somewhat low quality places are found  over the north-eastern Andes of Colombia, over mountain systems along BUC, MAC and PIS with clear sky fraction between 30-50\% during both periods of 2012. Southern Andes regions in the \textit{Nudo de los Pastos}, especially those in northern Ecuador display CSF values below 30\% from November to May and $\sim$70\% from June to October. Very low percentages of clear sky fraction below 20\% characterize central Andes of Colombia over the regions comprising \textit{Popay\'an} (POP) and BOG with the exception of the region centered at the high site \textit{Nevado del Huila} (75$^{\circ}$55'12"W, 2$^{\circ}$51'36"N) at 3700 meters with a mean of 35\% between December-February and June-July and 33\% for the other months.  

\begin{figure}
\centering
\begin{tabular}{cc}
 
    \includegraphics[trim=3.000cm 0.2cm 3cm 0.5cm,clip=true,width=0.55\textwidth]{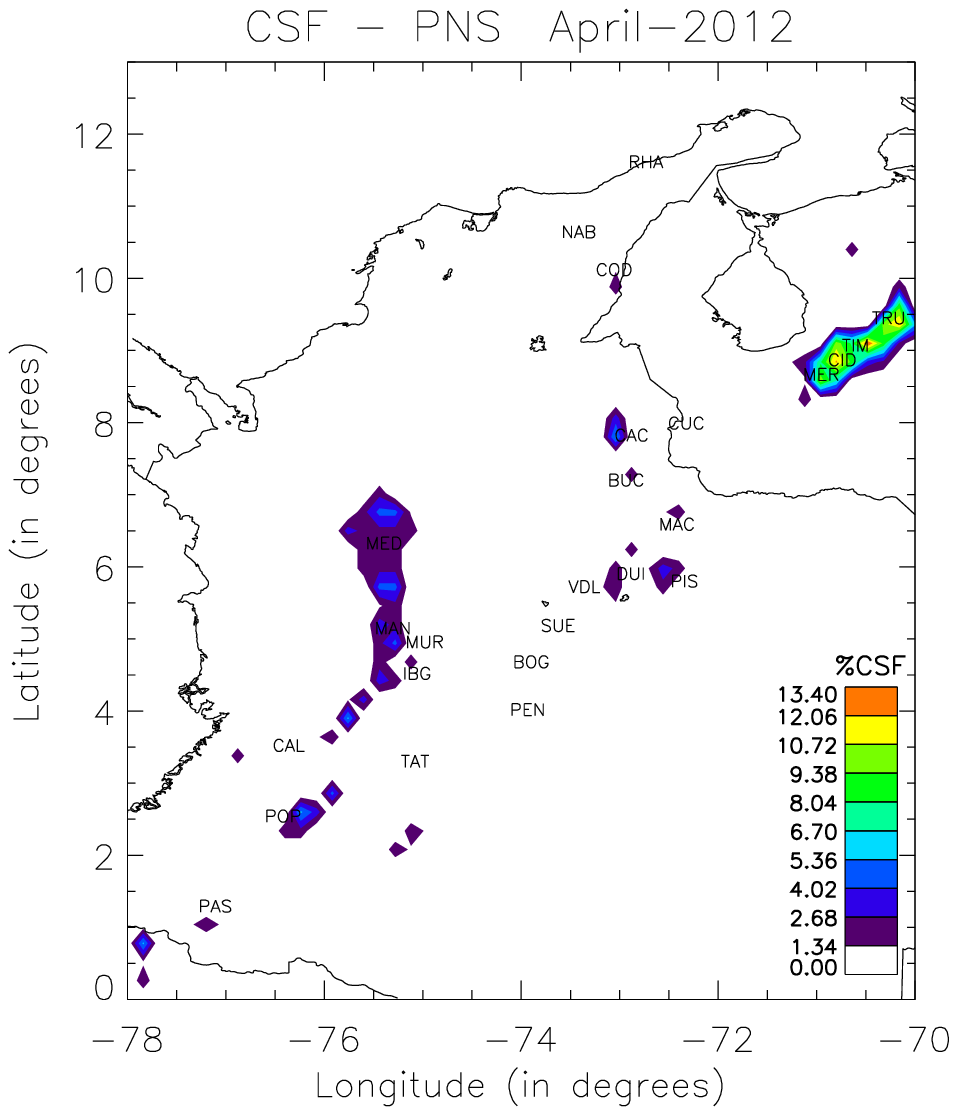}
    & \includegraphics[trim=3.000cm 0.2cm 3cm 0.5cm,clip=true,width=0.55\textwidth]{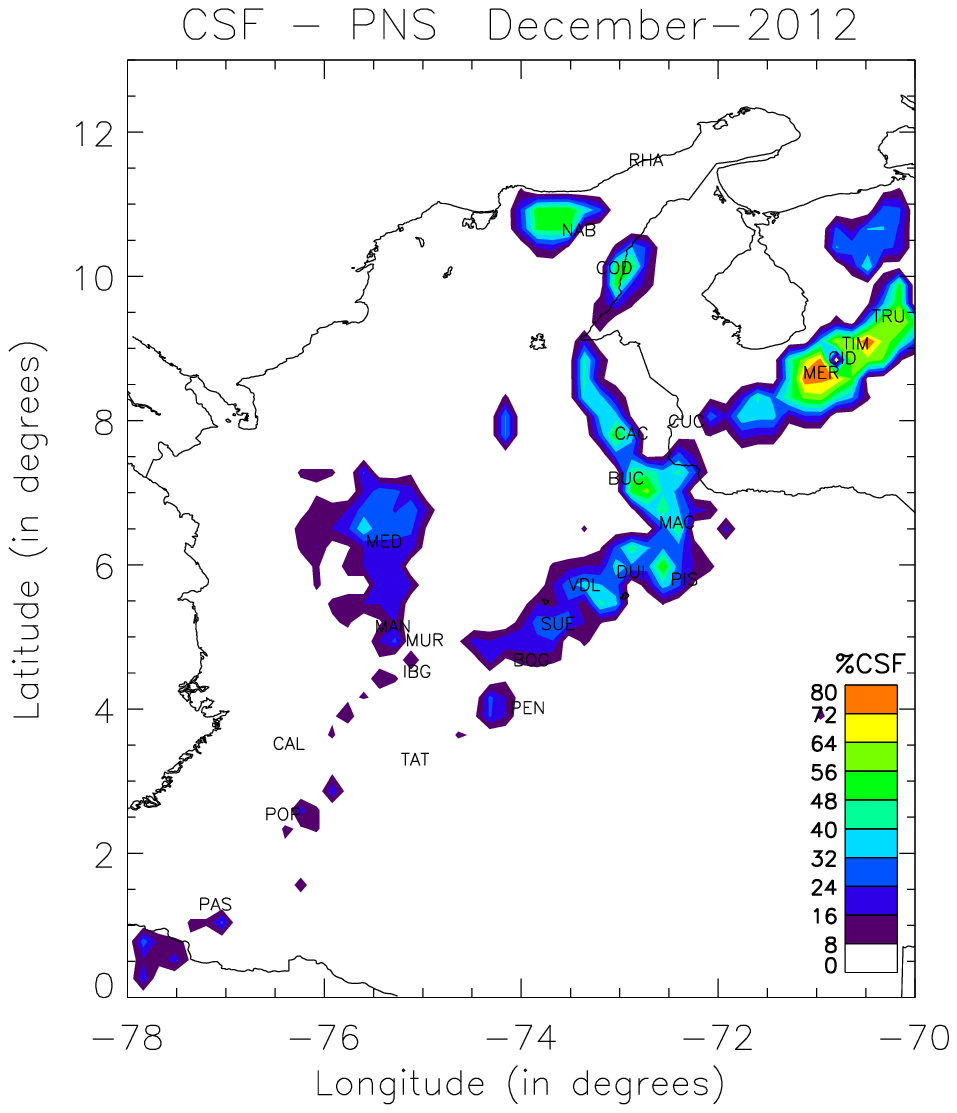} \\
\end{tabular}
\caption{Monthly clear sky fraction percentage for photometric nights (PNS) in the extended region for the year 2012  for April and December which are representative of the rainy and dry seasons, respectively.}\label{F11}
\end{figure}

Regarding monthly averages of CSF for PNS nights, the same regions were found  but with low percentages in comparison as shown in Figure \ref{F11}. The absence of spectroscopic nights in central Colombia from May to October confirms that November-February and June-July are marked by the largest CSF percentages while March-May and August-November are characterized by lower CSF values. We also observe such behaviour for other years. The existence of this bimodal behaviour with maxima during the dry season and minima  during the  rainy season is atributed, as we metioned already, to the double passage of the intertropical convergence zone for the extended region. Therefore we conducted yearly SNS and PNS simulations, bringing together the months December, January, February, June, July and August into one group that we call \textit{dry season} and March, April, May, September, October and November into another group denoted as \textit{rainy season}. 

SNS percentages for both seasons show maxima at the same locations than those observed for the year 2012 with seasonal differences of $\sim$20\% with the exception of low latitude locations close to the equator where slightly smaller differences are observed. Percentages for PNS simulations result 10\% lower in comparison to SNS nights but with identical CSF distribution over the extended region. 

Annual clear sky fraction averages obtained from our SNS simulations during rainy seasons for the years 2009 and 2012 are larger than for the other years. It seems to correlate with the changes in the Oceanic Ni\~no Index (ONI) reported by the climate prediction center NOAA\footnote{http://www.cpc.noaa.gov/products/analysis\_monitoring/ensostuff/ensoyears.shtml}.  Changes from positive to negative anomalies in April 2010 associated to a \textit{La Ni\~na} event increase the number and intensity of tropical storms in central Colombian Andes thus decreasing the clear sky percentages. The \textit{Nudo de los Pastos} and the \textit{Sierra Nevada de Santamarta} are not considerably affected by this \textit{El Ni\~no Southern Oscillation} (ENSO).

\section{Selected sites from five year CSF averages} 

\begin{figure}
\centering
\begin{tabular}{cc}
   \includegraphics[trim=3.000cm 0.2cm 3cm 0.5cm,clip=true,width=0.5\textwidth]{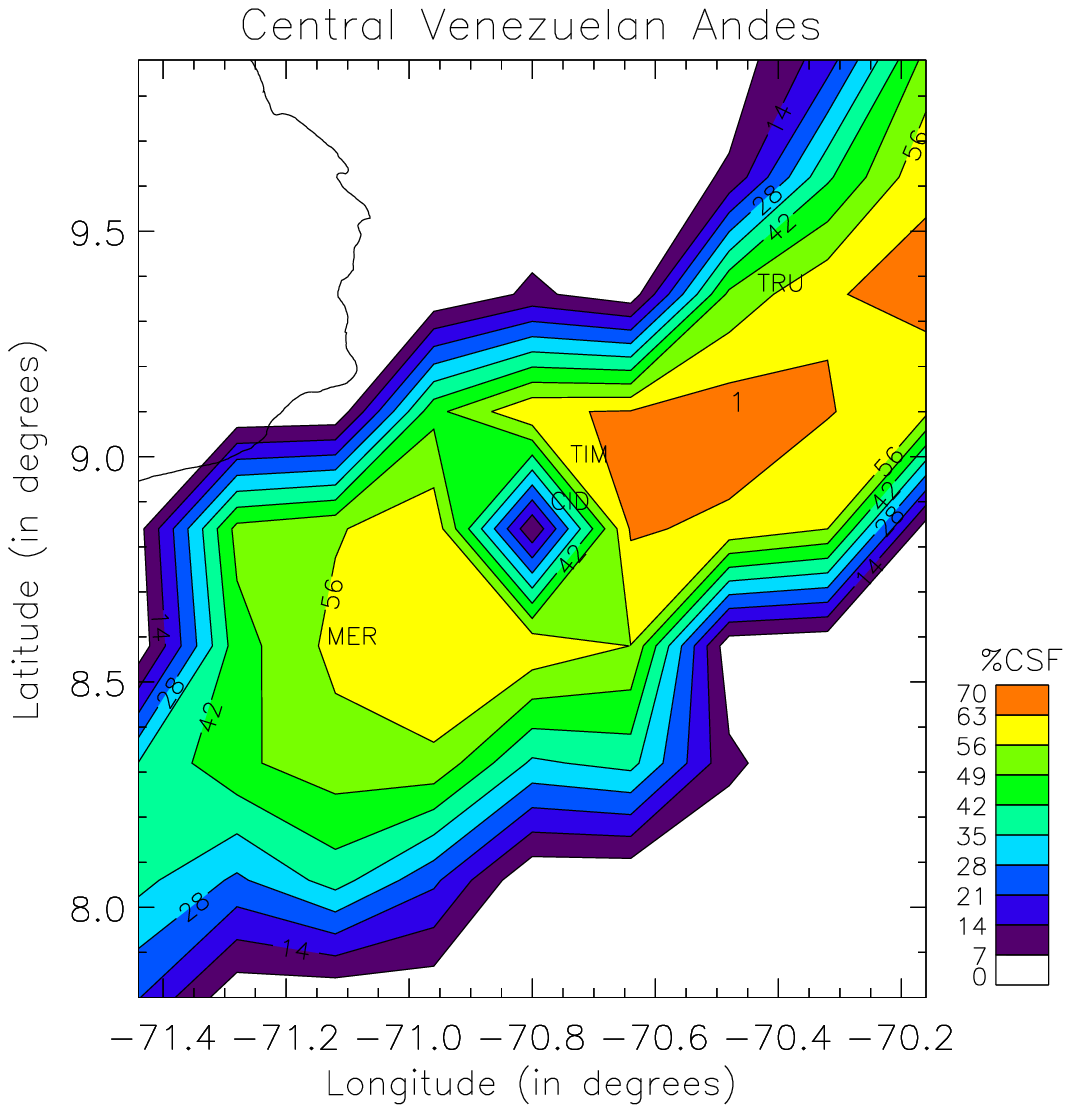} 
   & \includegraphics[trim=3.000cm 0.2cm 3cm 0.5cm,clip=true,width=0.5\textwidth]{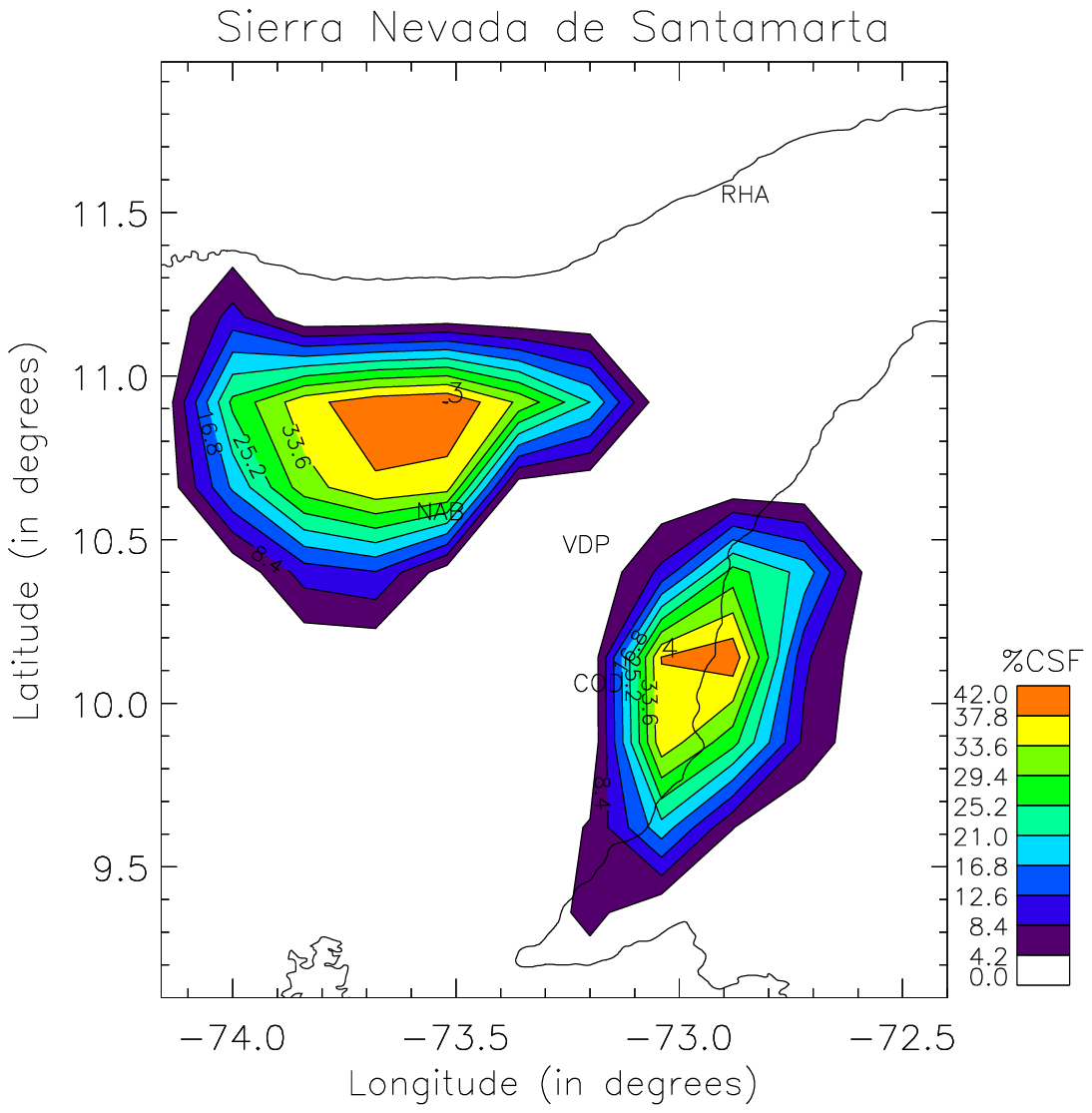}  \\
  \includegraphics[trim=3.000cm 0.2cm 3cm 0.5cm,clip=true,width=0.5\textwidth]{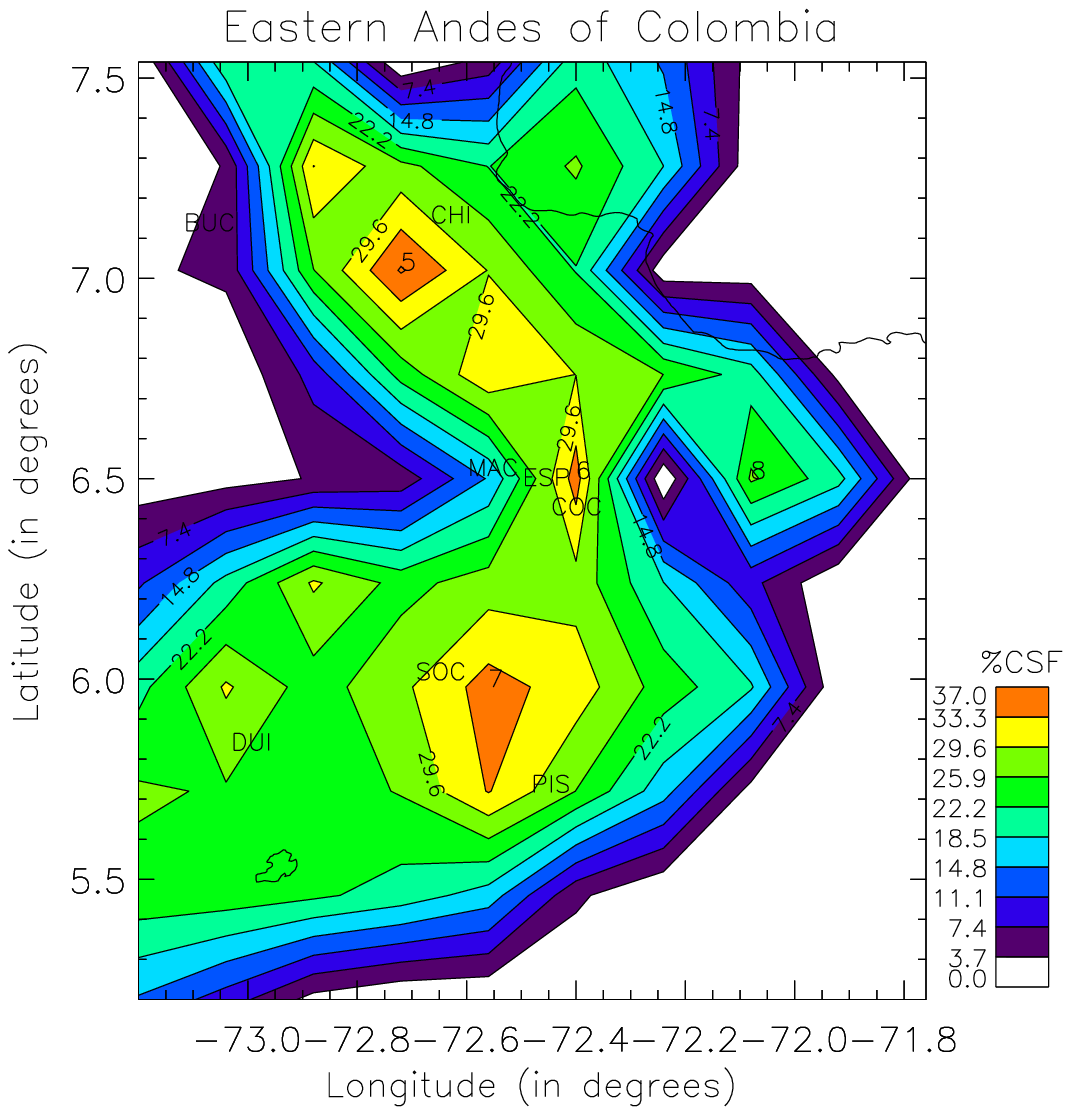} 
   & \includegraphics[trim=3.000cm 0.2cm 3cm 0.5cm,clip=true,width=0.5\textwidth]{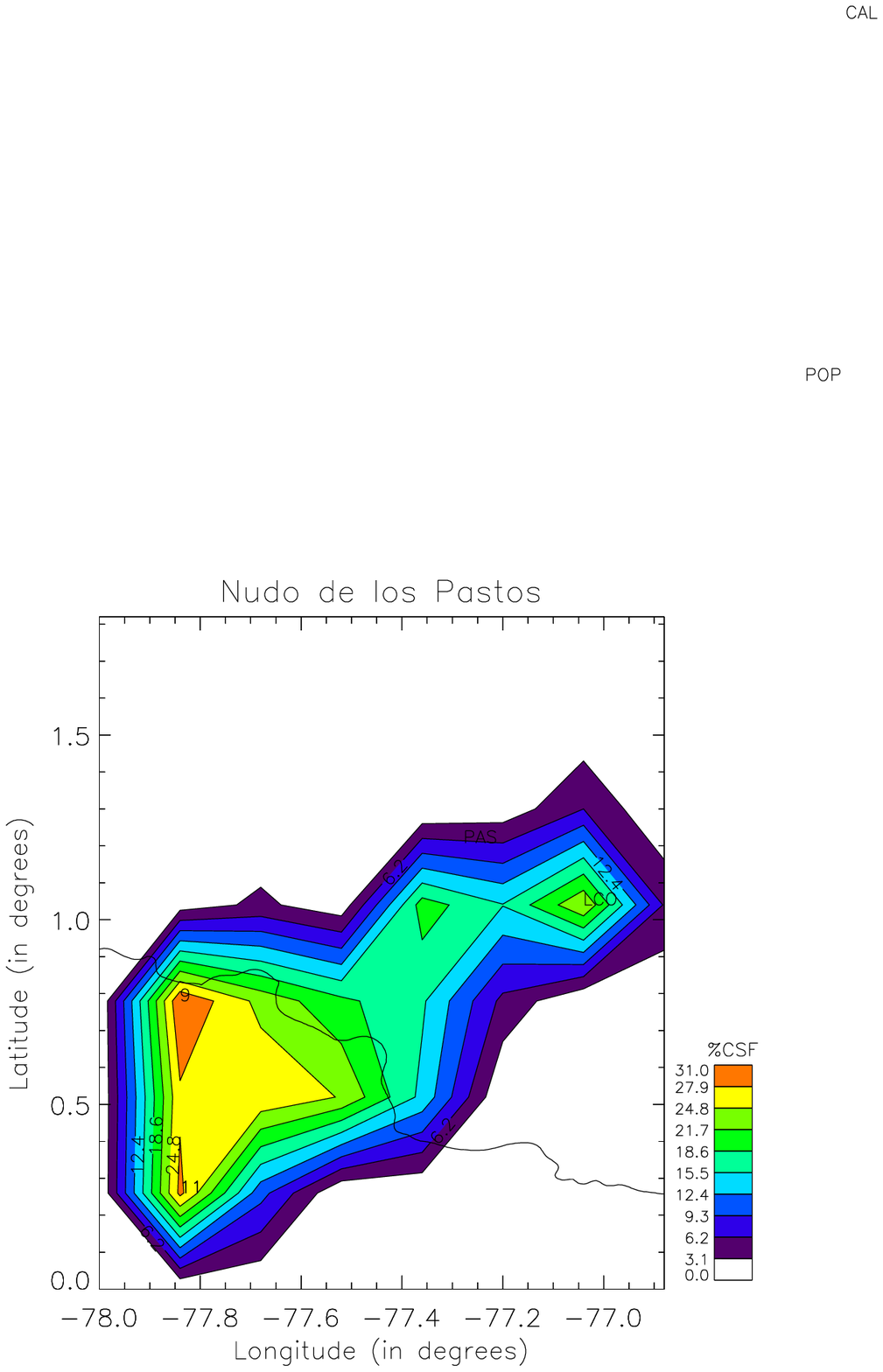}  \\
   \end{tabular}
\caption{Five year average of clear sky fraction for spectroscopic nights during the dry seasons in four regions containing the twelve selected sites in this study. }
\label{F16}
\end{figure}

For selection purposes we give preference to the simulations for spectroscopic nights during the dry seasons. We have selected twelve sites based on five year averages during the dry seasons that can be suitable for astronomical observations by selecting locations with CSF$\geq$30\%. Factors such as accesibility, security, service, local humidity, atomospheric \textit{seeing} were not considered and we need additional studies to confirm these locations as potential for astronomical sites. These twelve selected sites are distributed mostly along the following four zones: the central Andes of Venezuela comprising the states of \textit{M\'erida} and \textit{Trujillo}, the north east of the \textit{Sierra Nevada de Santamarta} around sites NAB and COD, the north eastern Andes of Colombia in the proximity of \textit{El Espino} (ESP) and a wide region on the southern Andes in \textit{Nudo de los Pastos} containing two locations in Ecuador. We show in Figure \ref{F16} the contours of clear sky fraction for these selected zones during the dry season with a spatial pixel resolution of 28$\times$18 km.  

Even though the selection criteria involves only spectroscopic nights during the dry seasons, we give a total number of clear nights per year at selected sites by adding the CSF five year averages for each season and then multiplying by a factor of 1.827. Tables \ref{tbl-2} and \ref{tbl-3} contain properties of the selected sites indicating the identifier, the nearest populated location, geographic coordinates of the site, altitude, yearly five year averages for the dry and rainy seasons, five year average of clear sky fraction, standard deviation and the total number of clear nights per year.  

\subsection{Central Venezuelan Andes}

The best places are situated over an extense area of the central Andes of Venezuela comprising latitudes from 8$^{\circ}$48'N to 9$^{\circ}$11'N and longitudes between 70$^{\circ}$17'W and 70$^{\circ}$42'W. The region contains a number of potential sites at high altitudes, in particular Site 1 (70$^{\circ}$28'48"W, 9$^{\circ}$5'60"N) where we obtain the maximum CSF in the extended region. Site 1 has an altitude of 3480 meters and it is located $\sim$35 km north-east from \textit{Timotes} (TIM) very close to the protected area \textit{Parque Nacional General Cruz Carrillo}. Five year average of clear sky fraction for spectroscopic nights during the dry seasons for this site is (67.0$\pm$10.7)\%  corresponding to 122.3 clear nights per year. The percentage during the rainy seasons is (53.5$\pm$12.2)\% leading to 97.7 clear nights per year. Subsequently we report a total number of (220$\pm$30) spectroscopic nights per year at Site 1. 

Regarding our reference Site 2 or CID, five year average of spectroscopic nights during dry seasons is (46.9$\pm$8.1)\% while for rainy seasons   it is (25.5$\pm$10.8)\% giving a mean of 72.4\% corresponding to  (132$\pm$25) total clear nights per year. Five year average for photometric nights at this site gives  (28.9$\pm$3.4)\% for dry seasons and (14.1$\pm$8.6)\% for rainy seasons giving a total of 43\%  which is in agreement with the log-books of the observatory as shown in Figure \ref{F9} even though seasonal differences were not considered for the validation of the algorithm as discussed in Section 6.  \\

\subsection{\textit{Sierra Nevada de Santamarta}}

This is a large system of mountains in the north of Colombia isolated from the Andes and located just 52 km from the Caribbean coast. It rises up to 5700 meters above sea level over a protected area of 17000 km$^2$ containing the national park \textit{Sierra Nevada de Santamarta}. Five year average of CSF is (42$\pm$10)\% over a broad region of 40 km to the north of NAB, specifically at Site 3 (73$^{\circ}$31'12"W, 10$^{\circ}$55'12"N) placed at an altitude of 3600 meters with 76.8 clear nights per year during the dry season and 49.3 nights per year during the rainy season. The total spectroscopic clear  nights at Site 3 is (126$\pm$34) in contrast with (72$\pm$16) clear nights per year with photometric quality. 

The steep gradient in elevation toward the south-east direction of NAB make the percentage of CSF lower along the plain regions on the valley of river \textit{Cesar} which separates the \textit{Sierra Nevada de Santamarta} from the \textit{Serran\'ia del Perij\'a},  an extension of the eastern Andean branch on the boundary between Colombia and Venezuela. Slightly lower percentages than those observed at Site 3 are obtained at Site 4 less than 10 km from the location COD. The site has (111$\pm$27) clear nights per year with spectroscopic quality and (66$\pm$18) photometric nights.   

 \subsection {Eastern Andes of Colombia}
 
This broad territory hosts a high number of \textit{P\'aramos}, dry forests, valleys and high snow peaks most ot them protected areas such as national parks \textit{El Cocuy} (COC) and \textit{Pisba} (PIS). The selected locations are within the wide region between the \textit{Sierra Nevada del Cocuy} and the north part of the canyon of the river \textit{Chicamocha}. The highlands of the sierra reach an elevation of more than 5000 meters including several snow peaks. Site 6 placed $\sim$50 km west from these high snow peaks, is the center of a region with five year average for the dry seasons of  (35$\pm$4)\%. Clear sky percentage toward the north west along the canyon reaches the maximum value of  (37.4$\pm$5.7)\% at Site 5 located in \textit{Cuchilla las Torrecitas}, next to \textit{Cordillera de Hoya Grande} in \textit{Norte de Santander} at an altitude of 4060 meters. For rainy seasons we obtain  (21.3$\pm$6.7)\% at this site giving a total of (107$\pm$16) clear nights per year. 

Site 7 located 20 km north-west of PIS with an altitude of 3600 meters belongs to the protected area \textit{Parque Nacional P\'aramo de Pisba}. Five year average is  (34.8$\pm$2.9)\% whereas  for rainy seasons Site 7 has  25\% giving a total of (99$\pm$16) clear nights per year. Another selected site in this region is Site 8, the closest selected location to the snow peaks, with (90$\pm$12) clear nights per year using only four years since the 2012 CSF average is below our detection threshold. Percentages of clear sky fraction for photometric nights for Sites 7 and 8 are indicated in Table  \ref{tbl-3}. 

Lower quality sites with CSF$\leq$15\% are clearly observed over the central branch of the Andes comprising locations \textit{Murillo} (MUR) and \textit{Medellin} (MED). Brieva (1985), using cloud cover data from visual inspection of the sky reports between 130 and 250 useful nights per year at site \textit{Ot\'un} (4$^{\circ}$46'33"N, 75$^{\circ}$24'53"W), located 25 km in the south west direction from MUR. We obtain (89$\pm$11) clear nights per year at site MUR (4$^{\circ}$59'19",75$^{\circ}$10'12").

\subsection{\textit{Nudo de los Pastos}}
 
This region consists of tropical humid forests where the Andes splits into the western and central branches. Sites 9 and 11 placed in north of Ecuador stand out as the best places in the region with (105$\pm$ 36) and (101$\pm$ 42) clear nights per year, respectively. With  (24.3$\pm$3.9)\% below our selection limit we include the Site 12 \textit{Laguna de la Cocha} (LCO) located at the Colombian side of the \textit{Nudo de los Pastos}.

Minimum seasonal differences are found at the sites in this region. Comparatively with the other sites, the percentage of clear sky during dry and rainy seasons is very similar in sites within the \textit{Nudo de los Pastos} confirming that the sky cover does not change significantly from one season to another at locations near the equator. We note from Tables \ref{tbl-2} and \ref{tbl-3} that the minimum differences between CSF values for dry and rainy seasons for all twelve sites are observed especially for the sites 9, 11 and 12 with differences of 4, 2 and 1\%. 

\begin{figure}
\centering
\begin{tabular}{cc}
 \includegraphics[width=13cm,height=8.5cm,angle=-90]{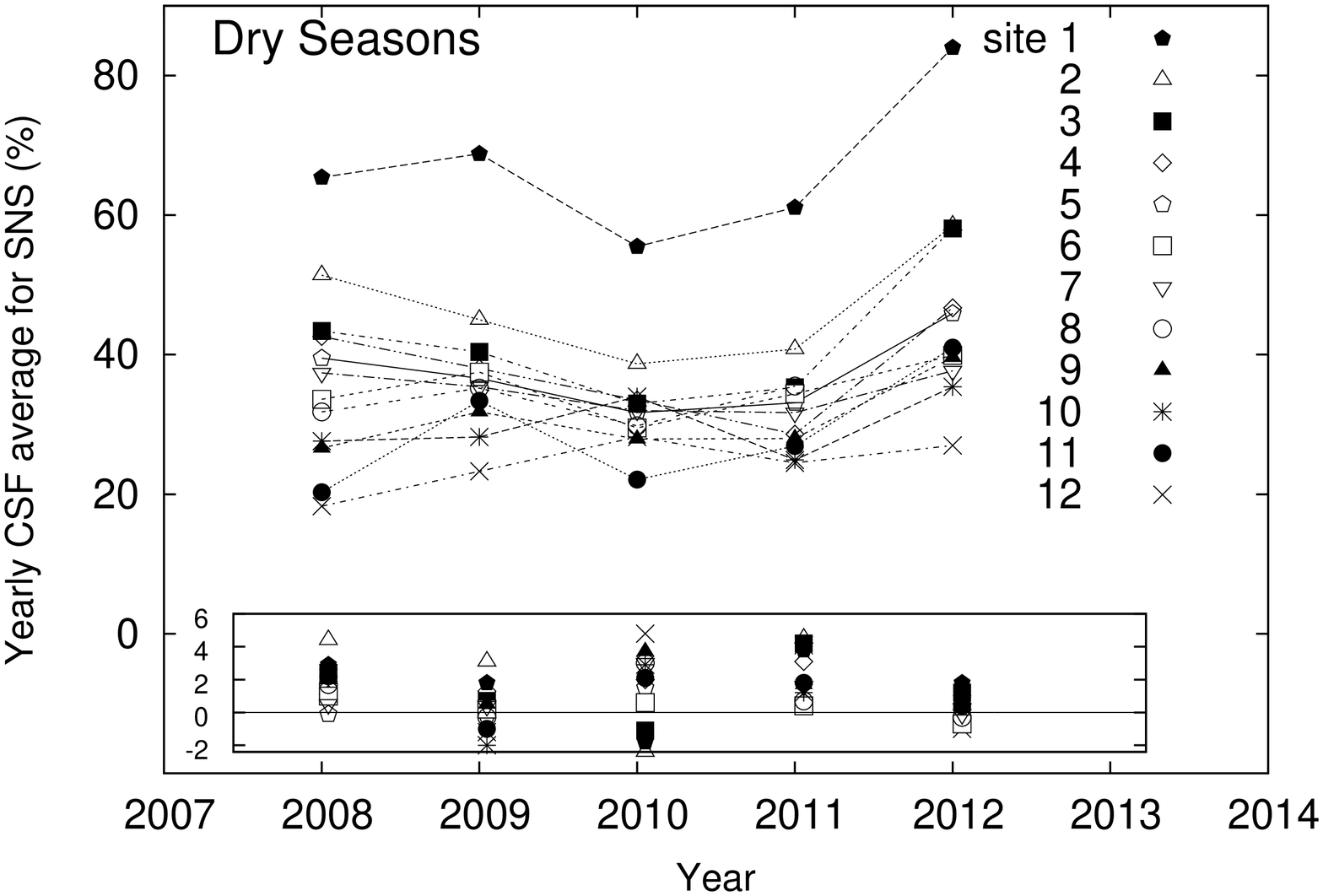}
   & \includegraphics[width=13cm,height=8.5cm,angle=-90]{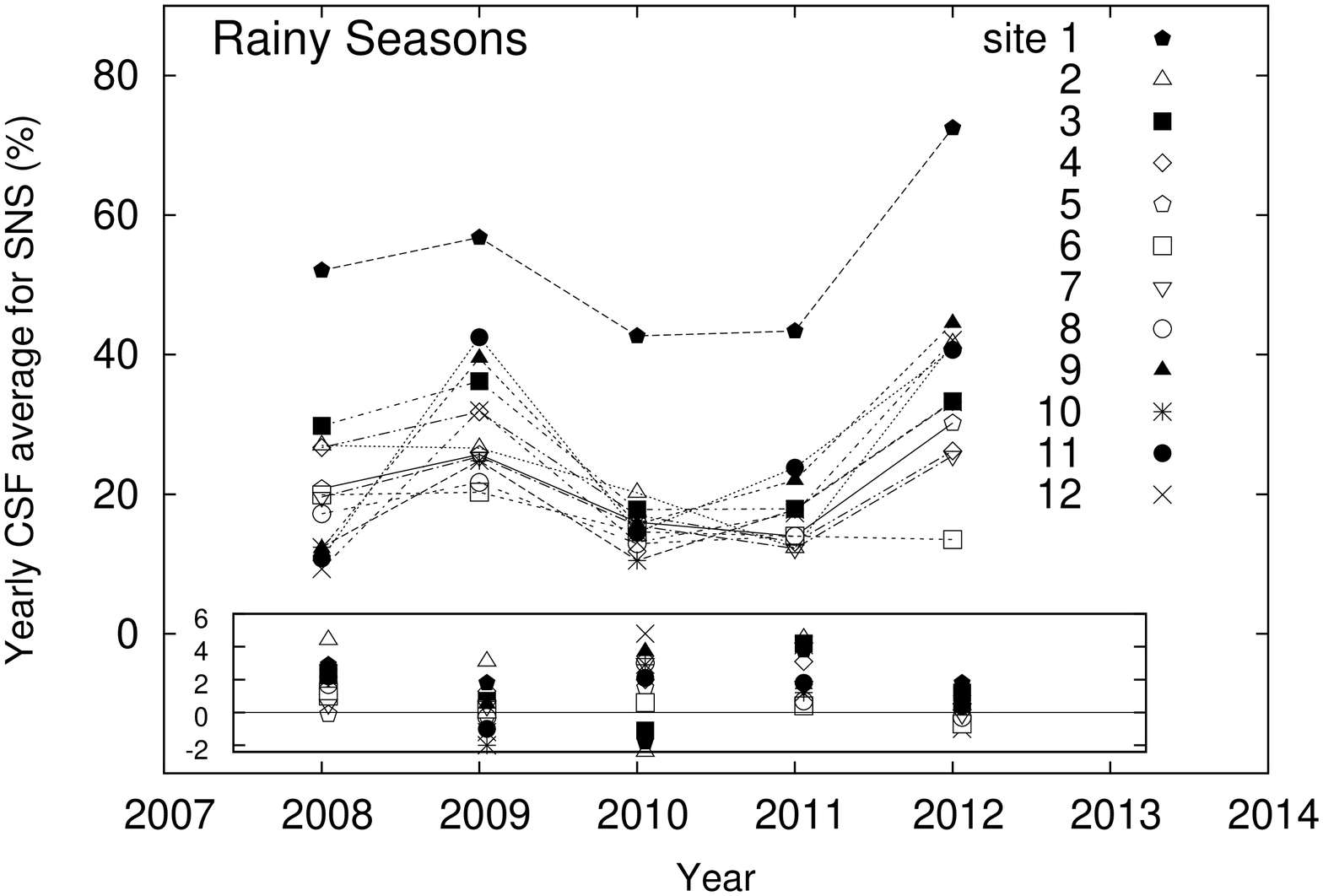} \\    
  
   \end{tabular}
\caption{Yearly clear sky fraction for spectroscopic nights for twelve selected sites for dry (left) and rainy (right) seasons of the years 2008-12. Dry Season comprises the months of December, January, February, June, July and August and Rainy Season the months of March, April, May, September, October and November. Interior plot corresponds to the Southern Oscillation Index (SOI) in the Ni\~no 3.4 region (-5S,5N;-120-170W) for the same years. Symbols of the monthly anomalies go along with the numbered sites.  }
\label{F17}
\end{figure}

Five year average for the selected sites  exhibit an extended minima in 2010-11 during rainy seasons as shown in Figure \ref{F17}. We found a possible correlation to \textit{La Ni\~na} episodes since positive values for the Southern Oscillation Index (SOI) are observed for these two years. Interior plots in Figure \ref{F17} show SOI values obtained from NOAA website for 2008-12. The symbols for the months go along with those for the sites. Positive values mostly for 2008, 2010 and 2011 are associated to intensification of storms over the Andes and then to low CSF values as observed.

\section{Conclusions}

In this work we describe a method to compute the clear sky fraction from GOES data based on comparisons with long-term records of temperature from weather stations and radiosondes, over a wide region of northern South-America comprising sites in Colombia, Venezuela and Ecuador.

We conducted a temporal analysis based on three daily values along five years doing a careful comparison of GOES12/13 temperatures in bands B3 and B4 with thresholds at different altitudes.  This approach enabled us to carefully determine the percentage of clear sky simultaneosly over the whole region. Similar perentages were obtained for all locations along the five years confirming the reliability of our methodology. 

Threshold monthly temperatures as a function of the altitude of the terrain were obtained for a small region of control through  Kriging interpolation of long-term records of temperature at weather stations. Monthly linear correlations between  thresholds and altitude were found and extrapolated to an extended region including \textit{Observatorio Nacional de Llano del Hato}. Threshold monthly temperatures at 8, 9 and 10 km along the upper atmosphere were also computed based on daily radiosonde data of two stations located at sea level and at the central Andes.  

The method differs from other cloud fraction algorithms that uses long-term records of temperature to obtain the CSF. Our approach gives more precise CSF values as long as we have used reliable reference temperatures at the surface and at the high troposphere.  Furthermore, our model can be easily extended to other tropical regions of astronomical interest. 

Nights are classified as spectroscopic or photometric based on the three CSF values during the night. The total number of spectroscopic and photometric nights per year at a fixed site are computed separately from five year averages for dry and rainy seasons. 

The best location found in this study is Site 1 (70$^{\circ}$28'48"W, 9$^{\circ}$5'60"N) in the central Venezuelan Andes with (220$\pm$42) spectroscopic clear nights per year and (157$\pm$24) photometric clear nights per year. Site 1 is placed at an altitude of 3480 meters and 35 km north-east from \textit{Timotes} within a protected area of Venezuela. Two lower quality locations were found in \textit{Sierra Nevada de Santamarta}, Site 3 (73$^{\circ}$31'12"W, 10$^{\circ}$55'12"N) and Site 4 (73$^{\circ}$2'24"W, 10$^{\circ}$8'24"N) with (126$\pm$34) and (111$\pm$27) spectroscopic clear nights per year, respectively. Site 3 stands  out as the best site identified in Colombia.

The eastern Andes of Colombia host two regions with more than 90 spectroscopic clear nights per year. The region centered at Site 5 (72$^{\circ}$43'12"W, 7$^{\circ}$1'12"N) in \textit{Norte de Santander} with (107$\pm$23) clear nights per year and the wide zone centered at \textit{El Espino} comprising \textit{Macaravita} and \textit{El Cocuy} in \textit{Boyac\'a} with (94$\pm$13) clear nights per year. The nearest locations to the equator in the \textit{Nudo de los Pastos} region are Site 9 ( 77$^{\circ}$50'24"W, 0$^{\circ}$46'48"N) and Site 11 (77$^{\circ}$50'24"W,  0$^{\circ}$15'36"N) with (105$\pm$36) and (101$\pm$42) clear nights per year, respectively, presenting minimum seasonal differences of number of clear nights as expected. Finally, CSF values below 25\% observed during the rainy seasons of the years 2010 and 2011 seems to be correlated to positive values of the Southern Oscillation Index. 

The selected sites obtained in this study may be considered as a first-step to decide the location of an astronomical observatory. However, site-survey is absolutely required before start any astronomical facility. Results of an observational campaign with the aim to characterize the sites identified in this study will be reported in a forthcoming paper. 

We would like to further point out that the value of our approach goes beyond the possibility of being useful for identification of potential sites in the region. Our approach allows us to quantify the contribution of the different layers of the atmosphere to the CSF above the site, in particular, the high troposphere and the surface. Although we have had to make a number of simplifying assumptions, the results are self-consistently and giv

\acknowledgments

This reasearch was supported by COLCIENCIAS and Universidad Nacional de Colombia through the project No. 110152129320. We thank \textit{Instituto de Hidrolog\'ia, Meteorolog\'ia y Estudios Ambientales de Colombia} (IDEAM) for providing meteorological data set. This publication makes use of data products from the OAN-Venezuela, which is operated by CIDA for the \textit{Ministerio para el Poder Popular para la Educaci\'on Universitaria, Ciencia y Tecnolog\'ia} of Venezuela. We thank the invaluable assistance of the observers and night assistants at the OAN-Venezuela and the support of the computer and technical department of CIDA in the creation of the database used in this contribution as validation 
of our method.

\clearpage



\clearpage


\begin{center}
\begin{table}
\begin{center}
\begin{tabular}{ccc}
{Month} & {$A$ ($^{\circ}$C/km) }& {$B$ ($^{\circ}$C)} \\
\hline
\hline
{JAN} & {-5.98$\pm$0.12} & {24.48$\pm$0.27}\\
{FEB} & {-6.01$\pm$0.13}& {24.42$\pm$0.29}\\
{MAR} & {-5.96$\pm$0.12} & {24.71$\pm$0.27}\\
{APR} & {-5.83$\pm$0.11} & {24.85$\pm$0.25}\\
{MAY} & {-5.71$\pm$0.11} & {24.60$\pm$0.26}\\
{JUN} & {-5.69$\pm$0.13} & {24.34$\pm$0.28}\\
{JUL} & {-5.93$\pm$0.15} & {24.47$\pm$0.33}\\
{AUG} & {-5.84$\pm$0.14} & {24.41$\pm$0.30}\\
{SEP} & {-5.79$\pm$0.12} & {24.35$\pm$0.26}\\
{OCT} & {-5.72$\pm$0.11} & {24.35$\pm$0.24}\\
{NOV} & {-5.66$\pm$0.11} & {24.28$\pm$0.26}\\
{DEC} & {-5.79$\pm$0.13} & {24.22$\pm$0.28}\\
\end{tabular}
\caption{Historical monthly lapse rates ($A$) and intercepts ($B$) obtained from linear regression of threshold temperatures (from Kriging) with the altitude of the weather station (i.e. $T_{0}(h) = A\times h + B$). Variations  of 0.1$^{\circ}$C/km in the lapse rate are significant for threshold computations whereas percentage errors in $B$ are of the order of 1\% and are considered insignificants.}\label{T1}
\end{center}
\end{table}
\end{center}

\clearpage

\clearpage

\begin{deluxetable}{cccccccccccccccccccc}
\tabletypesize{\scriptsize}
\rotate
\tablecaption{Percentages of clear sky fraction for spectroscopic nights (PNS) for selected sites.
Dry Season comprises the months of December, January, February, June, July and August and Rainy Season the months of March, April, May, September, October and November. \label{tbl-2}}
\tablewidth{0pt}
\tablehead{

\colhead{Site} & \colhead{ID} & \colhead{Nearest} &  \colhead{Lon.}& \colhead{Lat.} & \colhead{Alt.}& \multicolumn{2}{c}{2008}&\multicolumn{2}{c}{2009}& \multicolumn{2}{c}{2010} &\multicolumn{2}{c}{2011}&\multicolumn{2}{c}{2012}& \multicolumn{2}{c}{Mean CSF(\%)}   & No.nights\\

 & &Location & (W)&(N) &(m) &Dry& Rainy & Dry& Rainy & Dry & Rainy& Dry& Rainy& Dry & Rainy & Dry&  Rainy &per year
}
\startdata

1 & TIM  &Timotes & 70$^{\circ}$28'48" & 9$^{\circ}$5'60"  & 3480 & 65.4 & 52.1 & 68.8 & 56.8 & 55.5 & 42.7 &  61.1   &  43.4 & 84.0 & 72.5 &  67.0$\pm$10.7 & 53.5$\pm$12.2    &220$\pm$42 \tabularnewline
2 & CID & Cida &70$^{\circ}$52'12" & 8$^{\circ}$47'24"  & 3600 & 51.4 & 27.0 & 45.0& 26.6& 38.7 & 20.2 &  40.8    &  12.3 & 58.5 & 41.6 &   46.9$\pm$8.1&	25.5$\pm$10.8	 &132$\pm$34\tabularnewline
3 & NAB & Nabusimake &73$^{\circ}$31'12" & 10$^{\circ}$55'12" & 3300 & 43.4 & 29.8 & 40.4 & 36.2 & 33.0 & 17.8 &  35.3 &  17.9 & 58.1 & 33.3 &  42.0$\pm$9.9&	27.0$\pm$8.7	 &126$\pm$34 \tabularnewline
4 & COD &Codazzi & 73$^{\circ}$2'24" & 10$^{\circ}$8'24" & 2620     & 42.6 & 26.7 & 38.0 & 31.8 & 33.5 & 17.0 &  28.6  &  13.1 & 46.7 & 26.2 & 37.9$\pm$7.2 &23.0$\pm$7.7	&111$\pm$27\tabularnewline
5 & CHI & Chitag\'a&72$^{\circ}$43'12" & 7$^{\circ}$1'12"  & 4060 & 39.5 & 20.8 & 36.6 & 25.7 & 31.7 & 16.0 &  33.1  &  14.0 & 45.9 & 30.2 & 37.4$\pm$5.7&	21.3$\pm$6.7	 &107$\pm$23\tabularnewline
6 & ESP  &El Espino &72$^{\circ}$24'0"& 6$^{\circ}$30'0"  & 3620 & 33.6 & 19.9 & 37.5 & 20.3 & 29.5 & 14.6 &  34.4  &  14.0 & 39.9 & 13.5 &  35.0$\pm$4.0 &	16.5$\pm$3.3	&94$\pm$13 \tabularnewline
7 & PIS  &P. Pisba &72$^{\circ}$33'36" & 5$^{\circ}$58'48" & 3600 & 37.4 & 19.6 & 35.4 & 25.4 & 32.0 & 15.5 &  31.7  &  12.2 & 37.7 & 25.5 & 34.8$\pm2.9$&	19.6$\pm$5.9	 &99$\pm$16 \tabularnewline
8 & COC  & El Cocuy & 72$^{\circ}$4'48" & 6$^{\circ}$30'0"   & 3600     & 31.8 & 17.2 & 35.2 & 21.7 & 29.8 & 12.9 & 35.5   & 14.0   & ...   &  ...   &  33.1$\pm$2.8&16.5$\pm$3.9	&90$\pm$12\tabularnewline
9 & TUF  & P. Tuffino& 77$^{\circ}$50'24" & 0$^{\circ}$46'48"  & 3400    & 26.6 & 12.2& 31.8 & 39.5 & 27.9 & 15.7 & 28.0   & 22.0  & 39.6 & 44.5 &  30.8$\pm$5.3&26.8$\pm$14.4 &105$\pm$36\tabularnewline
10& HUI & N. Huila &75$^{\circ}$55'12" & 2$^{\circ}$51'36"  &  3700    & 27.6 & 12.4 & 28.2 & 24.8 & 34.0 & 10.5 & 24.9   & 17.8  & 35.4 & 33.2 & 30.0$\pm$4.5&19.7$\pm$9.4	 &91$\pm$25\tabularnewline
11 & IBA &Ibarra&  77$^{\circ}$50'24" & 0$^{\circ}$15'36"  & 3660     & 20.3 & 10.8 & 33.4 & 42.5 & 22.1 & 14.5 &  26.9  &  23.8 & 41.0 & 40.7 & 28.7$\pm$8.5&26.5$\pm$14.6	 &101$\pm$42\tabularnewline
12& LCO  & L. Cocha& 77$^{\circ}$2'24" & 1$^{\circ}$2'24"  &  3400    & 18.3 & 9.3 & 23.3 & 32.01 & 28.2 & 13.1 & 24.5   & 17.3 & 27.0 & 42.1 & 24.3$\pm$3.9&22.8$\pm$13.8    &86$\pm$32\tabularnewline
\enddata
\end{deluxetable}


\clearpage

\begin{deluxetable}{cccccccccccccccccccc}
\tabletypesize{\scriptsize}
\rotate
\tablecaption{Percentages of clear sky fraction for photometric nights for selected sites. Dry Season comprises the months of December, January, February, June, July and August and Rainy Season the months of March, April, May, September, October and November.\label{tbl-3}}
\tablewidth{0pt}
\tablehead{

\colhead{Site} & \colhead{ID} & \colhead{Nearest} &  \colhead{Lon.}& \colhead{Lat.} & \colhead{Alt.}& \multicolumn{2}{c}{2008}&\multicolumn{2}{c}{2009}& \multicolumn{2}{c}{2010} &\multicolumn{2}{c}{2011}&\multicolumn{2}{c}{2012}& \multicolumn{2}{c}{Mean CSF(\%)}   & No.nights\\

 & &Location & (W)&(N) &(m) &Dry& Rainy & Dry& Rainy & Dry & Rainy& Dry& Rainy& Dry & Rainy & Dry&  Rainy &per year
}
\startdata

1  & TIM  & Timotes    & 70$^{\circ}$28'48" & 9$^{\circ}$5'60"   & 3480 & 54.3 &  36.7 & 55.5  & 35.2 & 44.2   & 29.1 & 44.8    & 28.5  & 57.9   & 44.8& 51.3$\pm$6.4&34.9$\pm$6.6&	 157$\pm$24 \tabularnewline
2  & CID  & Cida       & 70$^{\circ}$52'12" & 8$^{\circ}$47'24" & 3600 & 29.6 &  14.1 & 29.3  & 13.1 & 25.8   & 13.1 & 25.7    & 3.1  & 34.0   & 27.3 & 28.9$\pm$3.4&14.1$\pm$8.6	 & 79$\pm$22\tabularnewline
3  & NAB  & Nabusimake & 73$^{\circ}$31'12" & 10$^{\circ}$55'12"  & 3300 & 31.8 &  17.9 & 24.2  & 16.9 & 24.3   & 8.7  & 24.2    & 7.1  & 30.5   & 11.4   &  27.0$\pm$3.8&12.4$\pm$4.8	&72$\pm$16\tabularnewline
4  & COD  & Codazzi    & 73$^{\circ}$2'24" & 10$^{\circ}$8'24" & 2620 & 29.5 &   15.8& 26.7  & 13.8 & 22.3   & 9.3  & 16.5    & 4.9  & 28.7   & 12.3& 24.7$\pm$5.4&	11.2$\pm$4.3     &66$\pm$18 \tabularnewline
5  & CHI  & Chitag\'a           & 72$^{\circ}$43'12" & 7$^{\circ}$1'12"  & 4060 & 27.6 &  10.8 & 23.2  & 11.4 & 20.7   & 8.2  & 20.7    & 6.7  & 29.3   & 10.9& 24.3$\pm$4&	9.6$\pm$2.0	 &62$\pm$11\tabularnewline
6  & ESP  & El Espino  & 72$^{\circ}$24'0"& 6$^{\circ}$30'0"  & 3620 & 22.0 &  7.6  & 23.4  &  8.1 & 23.2   & 8.8  & 24.0    & 8.1  & 26.2   &4.8  &  23.8$\pm$1.5&7.5$\pm$1.6	&57$\pm$6\tabularnewline
7  & PIS  & P. Pisba   & 72$^{\circ}$33'36" & 5$^{\circ}$58'48" & 3600 & 25.4 &  10.3 & 23.2  & 12.2 & 23.2   & 6.3  & 21.3    & 5.6  & 26.8   & 6.4 & 24.0$\pm$2.1&	8.2$\pm$2.9	&59$\pm$9 \tabularnewline
8 & COC   & El Cocuy   & 72$^{\circ}$4'48" & 6$^{\circ}$30'0"    &  3600    & 21.6 &  6.4  & 23.1  &  9.9 & 21.7   & 7.2  & 22.1    & 7.3  & ...  & ...  & 22.1$\pm$0.7&7.7$\pm$1.5   .1 &54$\pm$4 \tabularnewline
9  & TUF  & P. Tuffino & 77$^{\circ}$50'24" & 0$^{\circ}$46'48"  & 3400     & 6.8  &  6.5  & 18.6  & 17.5 & 15.6   & 8.5  & 15.2    &  9.6 & 25.1   & 28.6&  16.3$\pm$6.6&14.1$\pm$9.1&56$\pm$29\tabularnewline
10 & HUI & N. Huila    & 75$^{\circ}$55'12" & 2$^{\circ}$51'36"  &  3700    & 12.3 &  5.1  & 10.7  & 10.6 & 19.9   & 6.0  & 13.3    & 8.7  & 13.5   & 8.8 & 13.9$\pm$3.5&7.8$\pm$2.2  &40$\pm$11 \tabularnewline
11  & IBA & Ibarra     & 77$^{\circ}$50'24" & 0$^{\circ}$15'36"  & 3660 & 7.9  &  6.0  & 21.1  & 17.8 & 10.3   & 8.0  & 15.5    & 14.6 & 22.7   & 29.0& 15.5$\pm$6.5	&15.1$\pm$9.1   &56$\pm$29\tabularnewline
12 & LCO  & L. Cocha   &  77$^{\circ}$2'24" & 1$^{\circ}$2'24" &  3400    & 5.3  &  2.7  & 11.0  & 12.1 & 14.3   & 4.0  & 11.3    & 7.3  & 13.3   & 22.8&  11.0$\pm$3.5&9.8$\pm$8.1  &38$\pm$21\tabularnewline

\enddata
\end{deluxetable}





\begin{thebibliography}{}

\bibitem [Brieva (1985)]{b85} Brieva, E., RMxAA, 1985, 10, 397 
\bibitem [Cavazzani et al.(2012)]{cava12} Cavazzani, S., Ortolani, S. and Zitelli, V., 2012, \mnras, 419, 3081 
\bibitem [Cavazzani et al. (2013)]{cava13} Cavazzani, S. and Zitelli, V., 2013, \mnras, 429, 1849
\bibitem [Cavazzani et al. (2011)]{cava11} Cavazzani, S., Ortolani, S., Zitelli, V. and Maruccia, Y., 2011, \mnras, 411, 1271
\bibitem [Della Valle et al.(2010)]{delavalle} Della Valle, A., Maruccia,Y., Ortolani, S. and Zitelli, V., 2010, \mnras, 401, 1904-1916
\bibitem[Erasmus \& Sarazin(2002)]{es02} Erasmus, D.A. and Sarazin, M., 2002, ASP Conf. Ser., 266, Ed. J. Vernin, Z. Benkhaldoun, and C. Mu\~noz-Tu\~non, San Francisco, p.310
\bibitem[Erasmus et al.(2006)]{es06} Erasmus, D.A. and van Rooyen, R., 2006, Ground-based and Airborne Telescopes. Ed. Stepp, L. M., Proceedings of the SPIE, Volume 6267
\bibitem[Folkins et al.(2008)]{folkins} Folkins, I., Fueglistaler, S., Lesins, G. and Mitovski, T., 2008, J. Atmos. Sci., 65, 1019-1034 
\bibitem[Hidayat et al.(2012)]{H12} Hidayat, T., Mahasena, P., Dermawan, B., Hadi, P. and Premadi, P.W., 2012, \mnras, 427, 1903
\bibitem[Kneizys et al.(1996)]{k96} Kneizys, F.X., Abreu, L.W., Anderson, G. P., 1996, The MODTRAN 2/3 Report and LOWTRAN 7 Model, Atmos. Rad. Measurement
\bibitem [Mar\'in et al.(2015)]{marin15} Mar\'in, J. C., Pozo, D. and Cur\'e, M., 2015, A\&A, 573, A41
\bibitem[Poveda \& \'Alvarez(2011)]{Poveda} Poveda, G. \& \'Alvarez, D., 2011, Clim. Dyn., 36, 2233
\bibitem[Poveda \& Mesa(2000)]{PovedaMesa} Poveda, G. \& Mesa, O., 2000, Geophysical Research Letters, 27,11,1675
\bibitem[Rossow]{ros85}Rossow, W.R., Kinsella, E., Wolf, A. \& Gardner, L., 1985 International Satellite Cloud Climatology Project (ISCCP), WMO/TD-No. 58 (1985)
\bibitem[Soden \& Bretherton(1993)]{soden93} Soden, B. J. and Bretherton, F. P., 1993, J.Geophys.Res., 98, D9, 669-688 
\bibitem[Weinreb et al.1997]{wein97} Weinreb, M. P., Jamieson, M., Fulton, N., 1997, Applied Optics, 36, 6895
\end{thebibliography}
\end{document}